\documentclass[article,onefignum,onetabnum]{siamonline171218}

\usepackage{lipsum}
\usepackage{amsfonts}
\usepackage{graphicx}
\usepackage{epstopdf}
\usepackage{algorithmic}

\ifpdf
  \DeclareGraphicsExtensions{.eps,.pdf,.png,.jpg}
\else
  \DeclareGraphicsExtensions{.eps}
\fi

\usepackage{enumitem}
\setlist[enumerate]{leftmargin=.5in}
\setlist[itemize]{leftmargin=.5in}

\newsiamremark{remark}{Remark}
\newsiamremark{hypothesis}{Hypothesis}
\crefname{hypothesis}{Hypothesis}{Hypotheses}
\newsiamthm{claim}{Claim}

\headers{}{}

\title{Robust interpretation of electrochemical impedance spectra using numerical complex analysis}

\author{
Jithin D. George\thanks{NAISE, Northwestern University, Evanston, IL, USA; 
Mathematics and Computer Science Division, Argonne National Laboratory, Lemont, IL, USA 
(\email{jithingeorge2023@u.northwestern.edu}).}
\and
Willa Brenneis\thanks{Department of Chemical and Biological Engineering, Northwestern University, Evanston, IL, USA.}
\and
Jeffrey Richards\footnotemark[2]
\and
Jeffrey Lopez\footnotemark[2]
\and
Vinod K. Sangwan\thanks{Department of Materials Science and Engineering, Northwestern University, Evanston, IL, USA.}
\and
Dilara Meli\footnotemark[3]
\and
Heather Kurtz\footnotemark[3]
\and
Lincoln J. Lauhon\footnotemark[3]
\and
Jonathan Rivnay\thanks{Department of Biomedical Engineering, Northwestern University, Evanston, IL, USA}
\and
Mark C. Hersam \footnotemark[3]
\and
Maria K. Y. Chan\thanks{Center for Nanoscale Materials, Argonne National Laboratory, Lemont, IL, USA; 
NAISE, Northwestern University, Evanston, IL, USA 
(\email{mchan@anl.gov}).}
\and
Valerie Taylor \thanks{Mathematics and Computer Science Division, Argonne National Laboratory, Lemont, IL, USA;NAISE, Northwestern University, Evanston, IL, USA (\email{vtaylor@anl.gov}).}
}

\usepackage{amsopn}

\makeatletter
\newcommand*{\addFileDependency}[1]{
  \typeout{(#1)}
  \@addtofilelist{#1}
  \IfFileExists{#1}{}{\typeout{No file #1.}}
}
\makeatother

\newcommand*{\myexternaldocument}[1]{%
    \externaldocument{#1}%
    \addFileDependency{#1.tex}%
    \addFileDependency{#1.aux}%
}

\ifpdf
\hypersetup{
  pdftitle={An Example Article},
  pdfauthor={D. Doe, P. T. Frank, and J. E. Smith}
}
\fi

\myexternaldocument{ex_supplement}
\usepackage{booktabs}   
\usepackage{multirow}   
\usepackage{amsmath} 
\newcounter{affil}
\author{
Jithin D. George\thanks{NAISE, Northwestern University, Evanston, IL, USA; Mathematics and Computer Science Division, Argonne National Laboratory, Lemont, IL, USA. Email: \texttt{jithingeorge2023@u.northwestern.edu}}
\and
Willa Brenneis\thanks{Department of Chemical and Biological Engineering, Northwestern University, Evanston, IL, USA}
\and
Vinod K. Sangwan\thanks{Department of Materials Science and Engineering, Northwestern University, Evanston, IL, USA}
\and
Dilara Meli\footnotemark[3]
\and
Heather Kurtz\footnotemark[3]
\and
Jeffrey Richards\footnotemark[2]
\and
Lincoln J. Lauhon\footnotemark[3]
\and
Jonathan Rivnay\thanks{Department of Biomedical Engineering, Northwestern University, Evanston, IL, USA; Department of Materials Science and Engineering, Northwestern University, Evanston, IL, USA}
\and
Mark C. Hersam\thanks{Department of Materials Science and Engineering, Northwestern University, Evanston, IL, USA; Department of Chemistry, Northwestern University, Evanston, IL, USA; Department of Electrical and Computer Engineering, Northwestern University, Evanston, IL, USA}
\and
Jeffrey Lopez\footnotemark[2]
\and
Maria K. Y. Chan\thanks{Center for Nanoscale Materials, Argonne National Laboratory, Lemont, IL, USA; NAISE, Northwestern University, Evanston, IL, USA. Email: \texttt{mchan@anl.gov}}
\and
Valerie Taylor\thanks{Mathematics and Computer Science Division, Argonne National Laboratory, Lemont, IL, USA; NAISE, Northwestern University, Evanston, IL, USA. Email: \texttt{vtaylor@anl.gov}}
}

\begin{document}

\maketitle


\begin{abstract}
Electrochemical Impedance Spectroscopy (EIS) is a non-invasive technique widely used for understanding charge transfer and charge transport processes in electrochemical  systems and devices. Standard approaches for the interpretation of EIS data involve starting with  a hypothetical circuit model for the physical processes in the device based on experience/intuition, and then fitting the EIS data to this circuit model. This work explores a mathematical approach for extracting key characteristic features from EIS data by relying on fundamental principles of complex analysis. These characteristic features can ascertain the presence of inductors and constant phase elements (non-ideal capacitors) in circuit models and enable us to answer questions about the identifiability and uniqueness of equivalent circuit models. 
In certain scenarios such as models with only resistors and capacitors, we are able to enumerate all possible families of circuit models. Finally, we apply the mathematical framework presented here to real-world electrochemical systems and highlight results using impedance measurements from  a lithium-ion battery  coin cell.

\end{abstract}
\begin{keywords}
  EIS, equivalent circuits, impedance spectroscopy, chemical inductance, Warburg element
\end{keywords}

\begin{AMS}
 
\end{AMS}

\section{Introduction}

Electrochemical Impedance Spectroscopy (EIS) is a standard technique for characterizing electrochemical systems through their impedance response when a small alternating current or voltage signal is varied across their input. This impedance response is typically fitted to theoretical models, which allows for a broader understanding of fundamental dynamical processes involving charges, ions, radicals, chemical species, defects, etc. The theoretical models have conventionally been equivalent circuit models where electrochemical processes such as diffusion and charge transfer are aggregated and approximated using circuit analogs such as resistors, capacitors, and inductors \cite{lazanas2023electrochemical}. 

The general nature of the EIS technique and analysis makes it applicable to a wide variety of electrochemical systems and electronic/ionic/photochemical devices, including transistors, solar cells, batteries, fuel cells, photoelectrochemical cells, corrosion, chemical sensors, and biosensors \cite{magar2021electrochemical, choi2020modeling, lazanas2023electrochemical, cano2010use, hernandez2020electrochemical, grossi2017electrical, lu2022timescale, doi:10.1021/acs.energyfuels.4c05715, doi:10.1021/acsami.9b03884}. There are software libraries such as PyEIS \cite{knudsen2019pyeis} and Impedance \cite{Murbach2020} that EIS practitioners use for equivalent circuit modeling and analysis. Another emerging EIS analysis approach, Distribution of Relaxation Times (DRT) \cite{ciucci2019modeling, boukamp2018use, dierickx2020distribution, lu2022timescale} is also often used to identify the distinct time scales for different and often competing chemical processes. To accelerate and improve the robustness of EIS data analysis, there has been an interest in automating the analysis using machine learning methods \cite{bongiorno2022exploring, buteau2019analysis, doonyapisut2023analysis, liu2020gaussian, lu2022timescale, BABAEIYAZDI2021120116, ZHU2019113627, ZHAO2022140350, BUCHICCHIO2023128461}. Deep neural networks \cite{doonyapisut2023analysis, bongiorno2022exploring}, support vector machines \cite{ZHU2019113627}, AdaBoost, and Random Forest \cite{ZHAO2022140350} are some of the techniques employed to identify equivalent circuit models from the impedance dataset. Machine learning approaches have also been used to extract features such as the state of charge for batteries \cite{BUCHICCHIO2023128461, BABAEIYAZDI2021120116}, to improve DRT analysis \cite{ciucci2020gaussian, liu2020gaussian, maradesa2022probabilistic}, and to automate the fitting process itself \cite{buteau2019analysis}. While machine learning approaches show promise for large-scale automated discovery, a ``one size fits all'' approach leaves a lot of room for misinterpretation. The machine learning approaches need to be coupled with constraints from theory and expert knowledge of specific materials in order to narrow down the field of possible function approximations for sub-classes of material systems.

Despite the widespread applications of EIS analysis, it is far from a one-shot approach for drawing reliable conclusions about fundamental materials processes. Two major challenges arise in EIS:
\begin{itemize}
    \item \textbf{The assumed model(s) might not be the right one(s).}
    
 A common practice in EIS analysis is to assume a few candidate circuit models suggested by a domain expert based on the physical processes expected to occur in the material. Without careful input from a domain expert, the assumed model(s) can be unphysical \cite{orazem2024proper}. Restricting analysis to a narrow set of models can also lead to important physical processes being overlooked. At the same time, if a large set of models is used to fit the data, noise in the data can nudge the fitting pipeline toward a circuit that does not fully capture the underlying processes in the material.

    \item \textbf{The EIS data might have been collected ineffectively.}

    EIS measurements are typically collected at logarithmically distributed frequencies ranging from 0.01 Hz to 100 kHz \cite{lazanas2023electrochemical}. However, various physical processes/equivalent circuit elements are dominant in different frequency regimes. Measurements sampled in such a standard uniform manner may ignore key dynamic processes in the material systems. Various kinds of noise are also dominant in particular frequency regimes.
For example, when EIS is applied to materials under shear, the rotational velocity can become comparable to the frequency at certain shear rates. When this happens, small imperfections in the sample can cause significant variations in the impedance measurements \cite{doi:10.1073/pnas.2203470119, doi:10.1073/pnas.2403000121}. The measuring instruments also carry with them frequency-dependent noise, further reducing the reliability of the data.
\end{itemize}

These two challenges are interrelated. Since different systems require tailored frequency sampling strategies, improper sampling directly contributes to the uncertainty in model selection. Consequently, a robust framework for model identification can turn uncertainty in model selection into a guide for pinpointing frequency regions where additional measurements are most informative.
\begin{figure}

    \includegraphics[width=\textwidth]{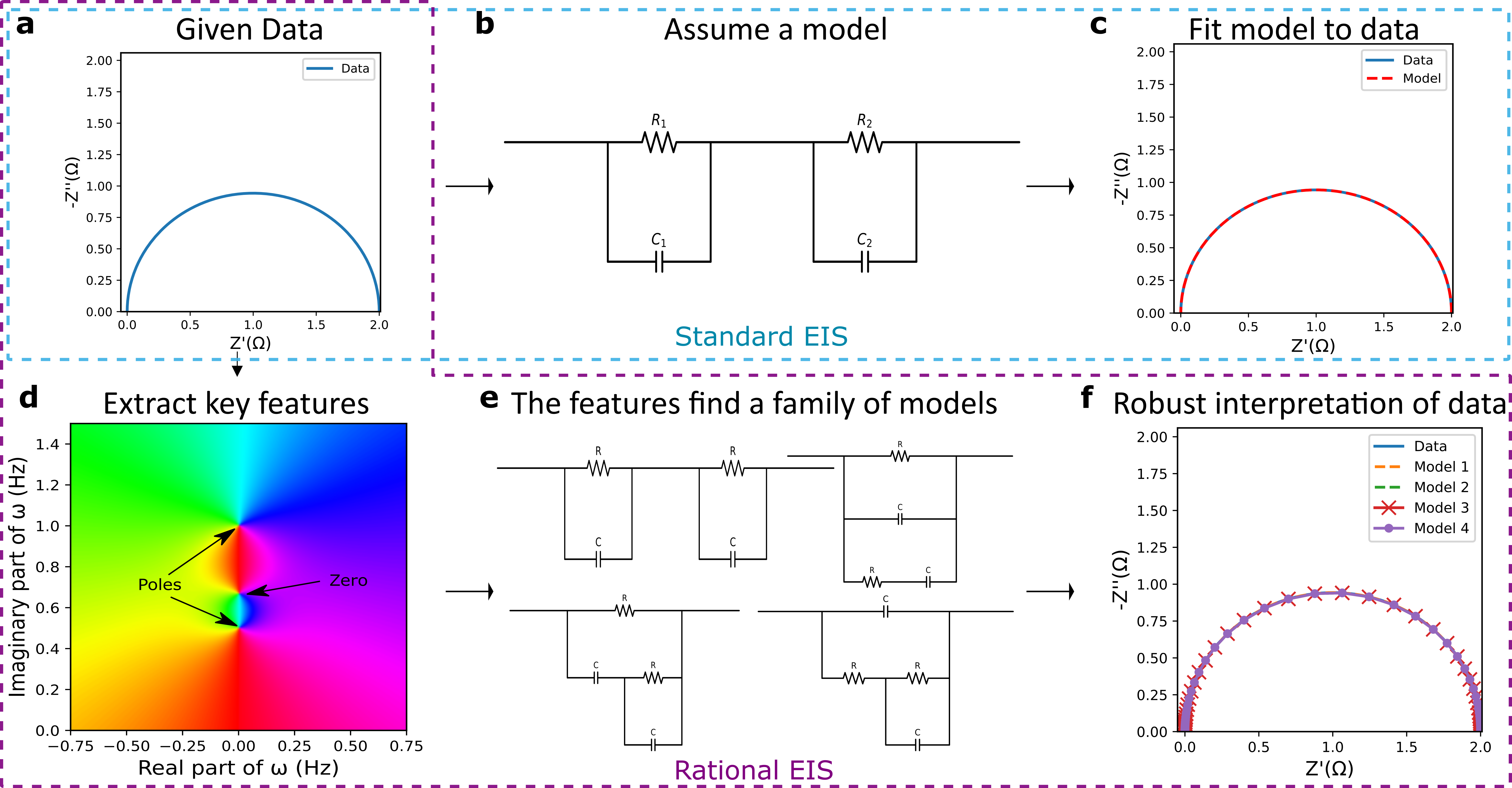}
    \caption{ (a, b, c) The standard approach to EIS analysis involves assuming a model given impedance spectra and fitting the model to the data to extract physical parameters. (a, d, e, f) Numerical complex analysis enables us to extract characteristic features directly from impedance without assuming a model. These features can be used to find families of models that share the same structure, allowing for reliable interpretation of data.}
    \label{fig:rationalEIS}
\end{figure}

The goal of the mathematical framework presented in this paper is to address these limitations of EIS using fundamental ideas from complex analysis. Impedance is a complex-valued function of frequency and  complex valued functions are characterized by their poles and zeros \cite{brown2009complex}. As a result, poles and zeros act as signatures of impedance datasets and valid models must exhibit similar signatures. Candidate models can be discovered by looking for circuits with particular pole-zero structure (See Figure \ref{fig:rationalEIS}). The pole-zero features correspond to relevant physical processes/circuit elements, which are dominant in particular frequency regimes. These features, obtained directly from data, can thus inform the experimental data collection and sampling for further validation and refinement of the set of models.

\begin{figure}
    \centering
    \includegraphics[width=\textwidth]{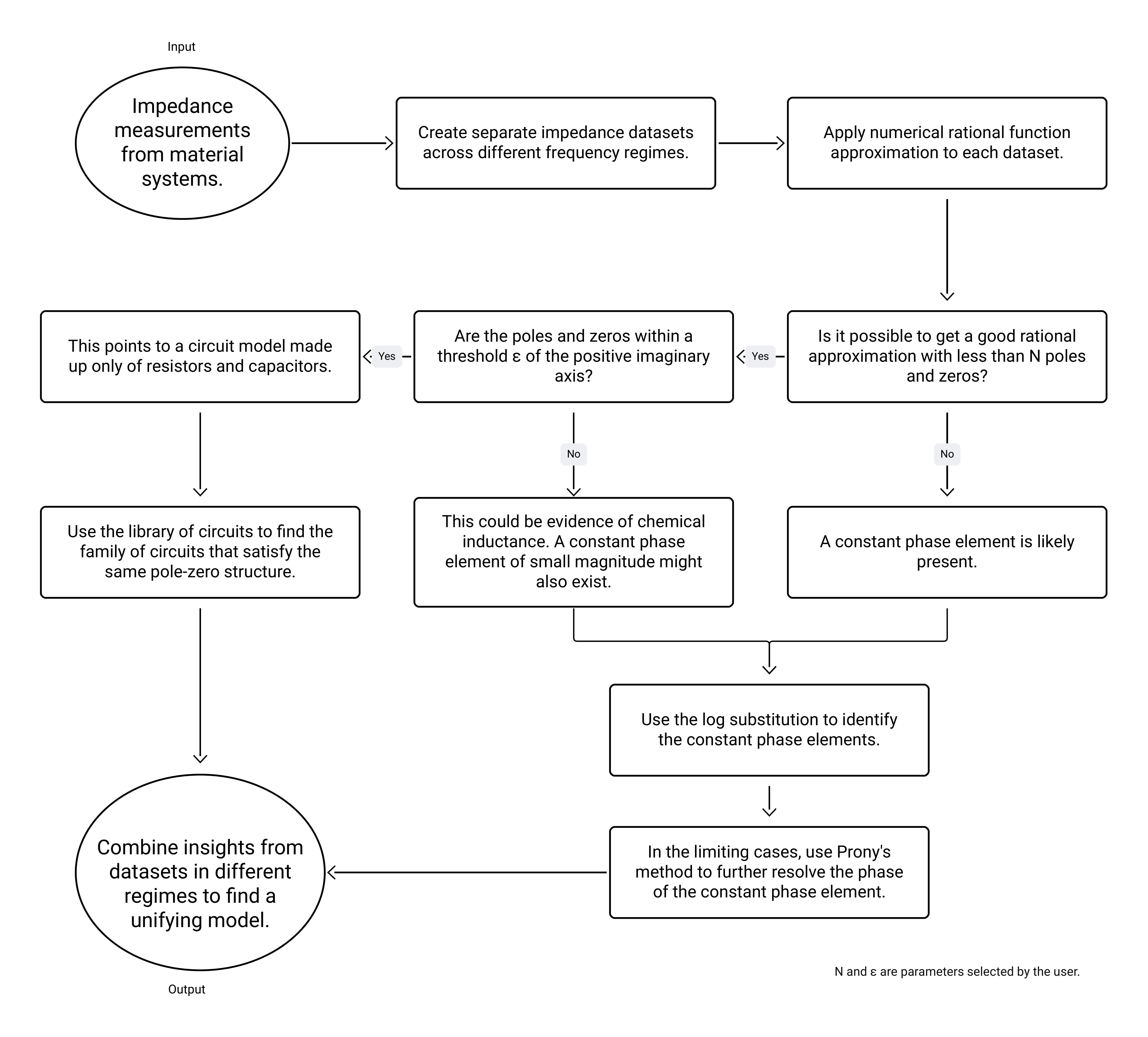}
    \caption{Schematic overview of the impedance analysis pipeline based on rational approximation. This model-agnostic framework begins with raw impedance measurements and uses rational approximation to extract features (poles and zeros) that uniquely characterize the system across various frequency regimes without requiring a predefined model. These features enable the discovery and identification of distinct capacitive and inductive behaviors, which can then be systematically combined to construct minimal interpretable circuit models.}
    \label{fig:pipeline}
\end{figure}

The paper is organized as follows: In Section \ref{sec:description}, we describe the recent advances in numerical analytic continuation that allow us to numerically extract characteristic features directly from impedance data and showcase a few examples to motivate the significance of these features. In Section \ref{sec:identifiability}, we mathematically explore whether it is possible to uniquely identify the underlying circuit from the impedance spectrum. This leads us to Section \ref{sec:properties}, where we identify key features of circuits with only resistors and capacitors that distinguish them from circuits with inductors or constant phase elements. Constant phase elements are non-ideal capacitors used to model the distributed capacitance that arises from surface roughness, porosity, or heterogeneity at electrode-electrolyte interfaces \cite{JORCIN20061473, doi:10.1021/jp512063f}. Section \ref{sec:CPE} explores mathematical approaches to detect and identify constant phase elements. We then apply our insights to experimental impedance datasets in Section \ref{sec:applications}. Finally, we discuss future prospects for improving the present approach and general applications. Figure \ref{fig:pipeline} illustrates how insights from our paper can be utilized in an analysis pipeline. 

\section{Description and motivation for  our approach}
\label{sec:description}

The task of relating the frequency-dependent impedance data to a theoretical model is a question of function approximation. Deep neural networks are universal function approximators \cite{liang2017deep, adcock2021gap, lu2021learning, hecht1987kolmogorov}, while classifiers choose from a small set of functions from a restrictive library \cite{doonyapisut2023analysis, kadam2020cnn}. The ideal function approximation should be general enough to capture the physically permissible limits and diversity of underlying processes but should also produce interpretable and realistic physical outputs.

In this paper, we use rational functions (i.e. functions of the form $f(z) = p(z)/q(z)$ where $p(z)$ and $q(z)$ are polynomials) to build a mathematical framework for modeling impedance. One reason is that the combined impedance of various constitutive processes typically ends up having a rational form, making it a natural choice to model the impedance function. Recent advances in numerical analytic continuation \cite{trefethen2023numerical,trefethen2020quantifying, nakatsukasa2018aaa} have enabled the extraction of extremely accurate rational function approximations of general functions directly from data. Most importantly, the impedance produced by typical circuit models, representing diverse constitutive physical processes, naturally assumes a rational form, making rational functions a compelling choice for modeling. 

Consider a circuit where a resistor with resistance $R_1$ in parallel with a capacitor with capacitance $C_1$, connected in series to another resistor with resistance $R_2$ in parallel with a capacitor with capacitance $C_2$. The impedance $Z(\omega)$ of this circuit is given by Equation \ref{eq:basic_poles}.
\begin{align}
\begin{split}
Z(\omega) = \frac{1}{\frac{1}{R_1}+C_1 i \omega}+\frac{1}{\frac{1}{R_2}+C_2 i \omega} = \frac{(C_1+C_2)(i\omega) + \frac{1}{R_1}+ \frac{1}{R_2}}{(\frac{1}{R_1}+C_1 i \omega)(\frac{1}{R_2}+C_2 i \omega)}
\end{split}
    \label{eq:basic_poles}
\end{align}
Since the right side of the Equation \ref{eq:basic_poles} is a rational function, it is evident that the impedance $Z(\omega)$ has a rational form. The root of the numerator in the rational representation, $\omega = \frac{1}{ C_1+C_2} \big(\frac{1}{R_1}+ \frac{1}{R_2}\big)i$, is the zero of the impedance function $Z(\omega)$. Similarly, the roots of the denominator $ \omega = \frac{1}{R_1 C_1}i$ and $\omega =\frac{1}{R_2 C_2}i $ are the poles of $Z(\omega)$.

 \begin{figure}[t]
    \centering
    \includegraphics[width=0.75\textwidth]{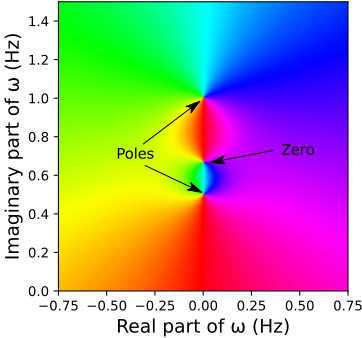}
    \caption{Phase portrait of the rational approximation to $Z(\omega)$ in Equation \ref{eq:basicrat_values}. The color indicates the phase of $Z(\omega)$ at each $\omega$ in the complex plane. A phase of zero corresponds to red, and as the phase increases, the color changes to yellow, green, blue, and then violet. Poles and zeros can be visually identified as points around which the phase changes from $0$ to $2 \pi$.}
    \label{fig:circuit_basic}
\end{figure}
The mathematical algorithm at the core of this paper, the Adaptive Antoulas-Anderson (AAA) algorithm \cite{nakatsukasa2018aaa}, enables us to extract a rational function representation and subsequently the poles and zeros of the impedance response directly from the numerical data. The AAA  algorithm and its variants \cite{driscoll2024aaa, wilber2022data, nakatsukasa2018aaa, baddoo2020lightning} have been employed for ``near-best'' numerical rational function approximation in a variety of applications such as nonlinear eigenvalue problems \cite{guttel2022robust}, efficient partial differential equation, fluid flow solvers \cite{gopal2019solving, baddoo2020lightning,  costa2023aaa, xue2024computation, nakatsukasa2021reciprocal}, model reduction \cite{yu2024leveraging, aumann2022automatic}, signal processing \cite{wilber2022data, derevianko2023esprit, betz2024efficient}, microwaves tubes \cite{xue2023rational}, magnet excited fields \cite{costa2024modelling}, wildfire modeling \cite{HARRIS2025106303}, and modeling relativistic quasiparticles \cite{kehry2023robust}.

The numerical rational approximation to Equation \ref{eq:basic_poles} produced by the AAA algorithm has the same  poles and zeros (within machine precision) as the circuit. Setting $C_1 = 1F, C_2 = 2 F, R_1 = R_2 = 1 \Omega$, we get
\begin{align}
Z(\omega) = \frac{1}{1+ i \omega}+\frac{1}{1 +2 i \omega} =- \frac{3i(\omega - \frac{2i}{3})}{2(\omega - i)( \omega - \frac{i}{2})}
\label{eq:basicrat_values}
\end{align}
This numerical rational approximation, constructed from Equation \ref{eq:basicrat_values} with only 30 data points, can be visualized through a phase portrait \cite{wegert2012visual}, shown in Figure \ref{fig:circuit_basic}.

A complex function of frequency $\omega$ (Hz), $Z(\omega)$ can be represented as $Z(\omega) = A(\omega) e^{i \phi(\omega)}$ where the real-valued amplitude $A(\omega)$ and the real-valued phase $\phi(\omega)$ are themselves functions of the input frequency $\omega$. A phase portrait visualizes the phase $\phi(\omega)$ of the function $Z(\omega)$ at every frequency $\omega$ value in the complex plane through colored pixels. A phase of zero corresponds to red, and as the phase increases, the color changes to yellow, green, blue, and then violet. Poles and zeros are characterized by phase changes as one circles around them. For a zero, the phase increases as we go in a counter-clockwise direction around it whereas the phase decreases for a pole. Figure \ref{fig:circuit_basic} shows that the rational approximation recovers the theoretical poles at $z = i$ and $z = \frac{i}{2}$ and the zero at $z = \frac{2i}{3}$.

Impedance measurements are typically visualized using Nyquist plots \cite{lazanas2023electrochemical}. A Nyquist plot displays the impedance data collected at particular frequencies in a feature space where the x-axis is the real component of the impedance and the y-axis is the negative of the imaginary component. Rational approximation, on the other hand, enables the impedance to be defined at all possible complex values of the input frequency $\omega$, not just at the collected frequency values. A phase portrait of the rational approximation visualizes the phase of impedance over the entirety of the complex plane. This allows us to compress the information contained in a Nyquist plot into a few key characteristic features, the poles and zeros of the impedance, which typically lie in the complex plane, away from the real line where the data is collected.
\begin{figure}
    \centering
\includegraphics[width=\textwidth]{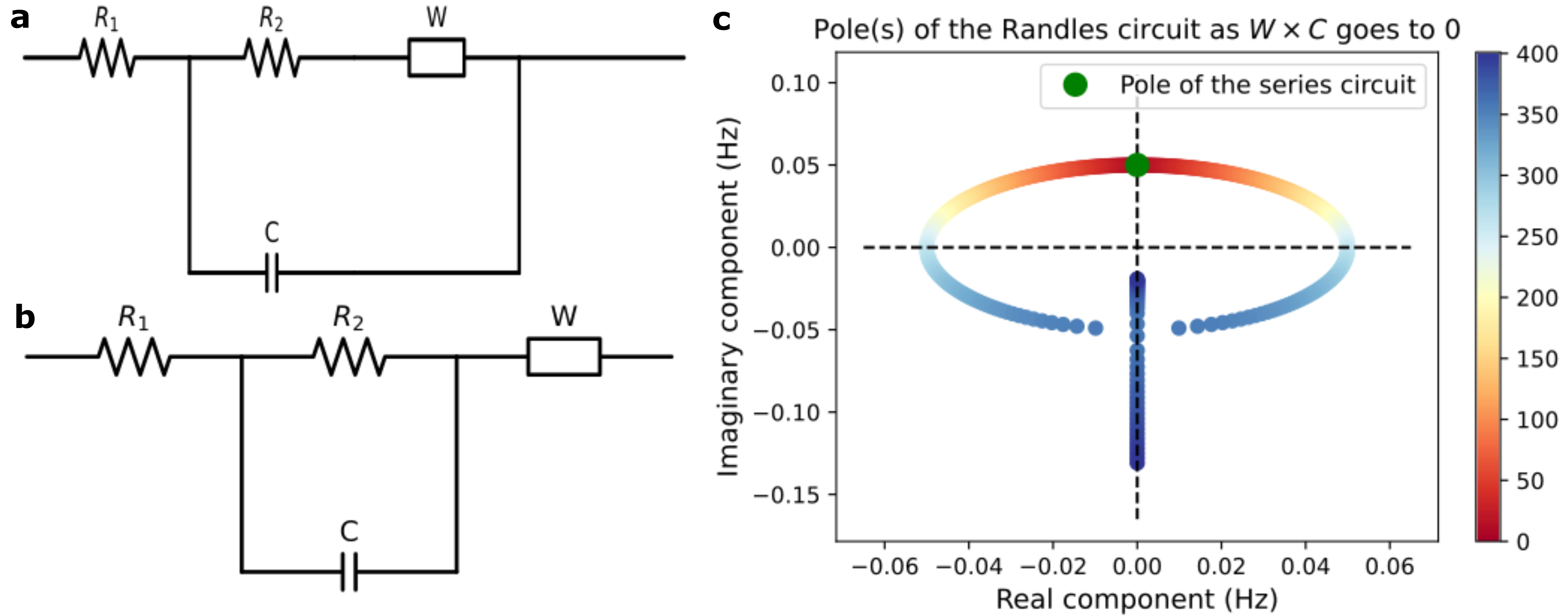}
    \caption{The two different circuits discussed in \cite{orazem2024proper}: a) The Randles circuit b) A circuit with the Warburg element as a standalone series component.  c) The pole of the circuit with the Warburg element as a standalone series component is compared in with corresponding poles of the Randles circuit. }
\label{fig:properwarburg}
\end{figure}

The poles and zeros extracted by the AAA algorithm give us significant insight into the underlying system. Two analytic functions \cite{brown2009complex} that share the same poles and zeros can be shown to be the same function, differing only by a constant multiplicative factor. To demonstrate the uniqueness of poles and zeros in impedance functions, we consider the two circuits in Figure \ref{fig:properwarburg} from ref. \cite{orazem2024proper}. In \cite{orazem2024proper}, the authors discuss how the two circuits, a Randles circuit and a circuit where the Warburg element is a standalone series component, have been used interchangeably and demonstrates flawed equivalence of two perceived equivalent circuit models through analysis of the underlying reaction kinetics. The impedance formulae of two circuits are shown in Equation \ref{eq:orazem}.

\begin{align}
\begin{split}
    Z_{Randles} &= R_1+\frac{R_2 + \frac{W}{\sqrt{i \omega}}}{1+i \omega C \bigg(R_2+\frac{W}{\sqrt{i \omega}} \bigg)}\\
    Z_{Series} &= R_1+\frac{R_2}{1+i \omega C R_2} +\frac{W}{\sqrt{i \omega}}
\end{split}
    \label{eq:orazem}
\end{align}

 In ref. \cite{orazem2024proper}, the authors subtract the two impedance formulae and state that the two circuits are equivalent as $W \times C$ goes to 0. The poles of the circuits, however, display very different behavior even as $W \times C$ goes to 0. The two circuits share the pole at 0. Comparing the other poles, the wildly different behavior of the circuits is evident. The pole of the series circuit always lies on the positive imaginary axis. The pole of the Randles circuit lies either on the negative imaginary axis or in the complex plane as a conjugate pair and only touches the positive imaginary axis in the limiting case of $W \times C =0$ (See Figure \ref{fig:properwarburg}). The poles of these two circuits, thus, enable us to quantify their wildly different behavior even when $W \times C \ll 1$.

\begin{figure}[h]
    \centering
    \includegraphics[width=\textwidth]{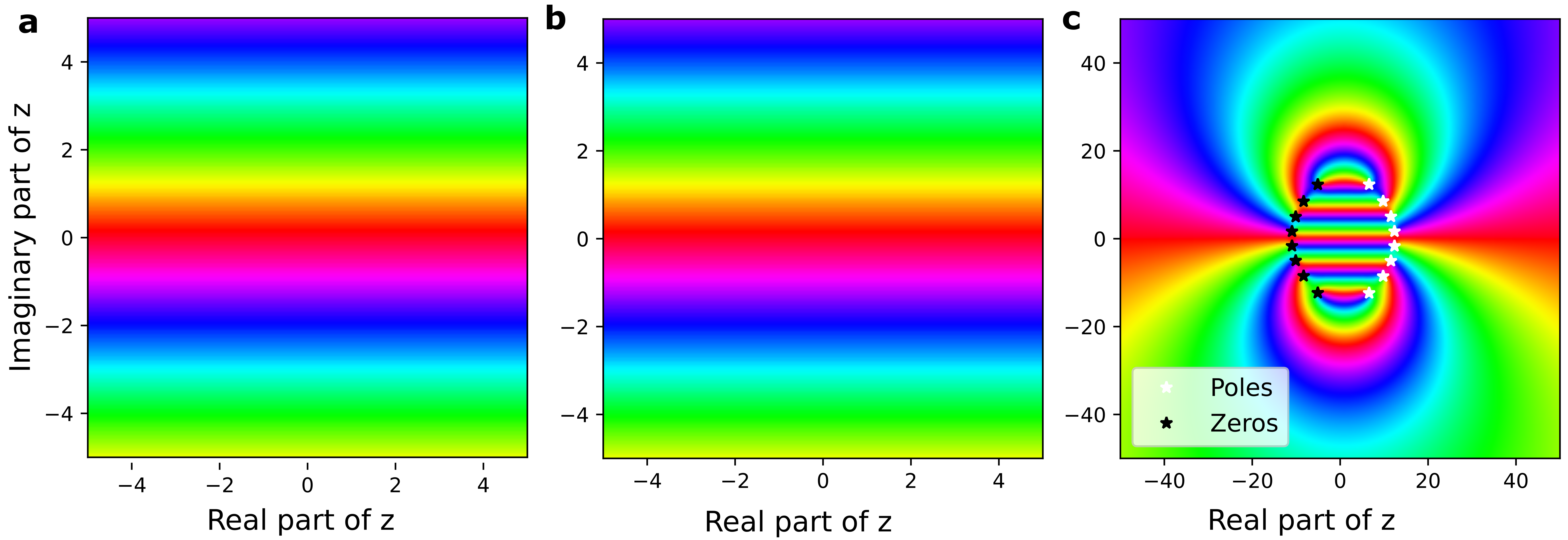}
    \caption{a) Phase portrait of the exponential function $\exp(z)$. b) Phase portrait of the rational approximation to the exponential function constructed using 40 $z$ values between 0 and 5. Although the exponential is not a finite rational function, the approximation closely matches its phase behavior within the box [-5,5] . c) Zoomed-out view of the phase portrait of the rational approximation.  The rational approximation places the poles and zeros at the locations highlighted here in order to ensure accurate representation of the exponential function within the unit box [-5,5].}
    \label{fig:pp_exp}
\end{figure}

\begin{figure}
    \centering
    \includegraphics[width=\textwidth]{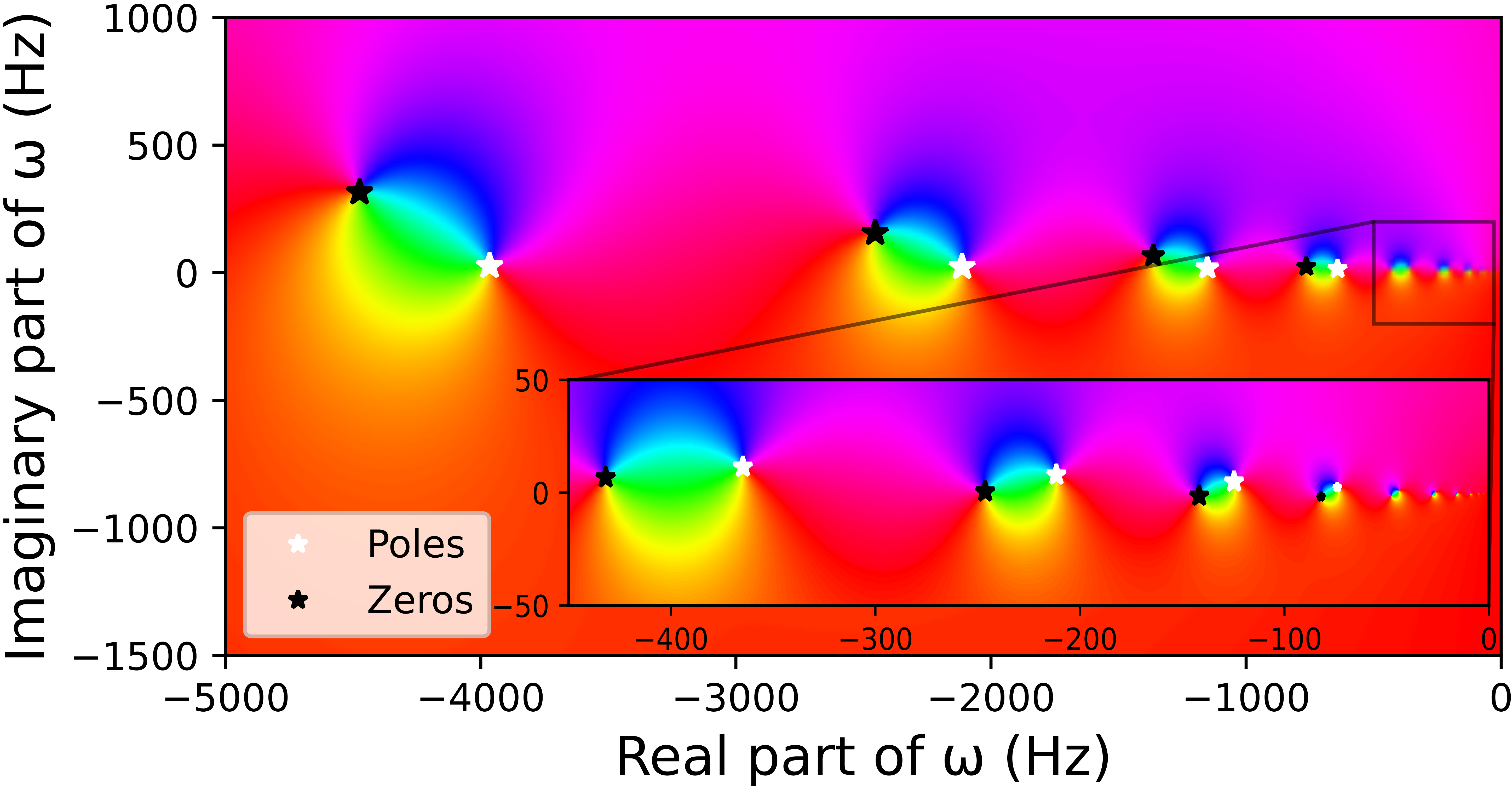}
    \caption{Rational function approximation enables us to check for constant phase elements and Warburg elements by looking for branch cuts in the phase portraits. Branch cuts appear as seemingly infinite series of poles and zeros in rational approximations.}
    \label{fig:branch_cut}
\end{figure}
The AAA algorithm also provides excellent rational approximation to functions that are not naturally in a rational form like the exponential function approximation in Figure \ref{fig:pp_exp}. This is particularly helpful in the case of circuits with constant phase elements and Warburg elements since they result in a special class of singularities known as branch cuts which are curves in the complex plane that separate different branches of multi-valued functions  \cite{brown2009complex}.

Consider a simple circuit with a constant phase element with impedance given by 
\begin{align}
    Z(\omega) = R_1 + \frac{1}{\frac{1}{R_2}+ A (i \omega)^\alpha}
    \label{eq:simple_CPE}
\end{align}

The presence of a constant phase element (an imperfect capacitor where $\alpha \neq 1$) fundamentally affects the rational function representation of the impedance. Due to the non-integer power $\alpha$,  Equation \ref{eq:simple_CPE} cannot be represented by a finite number of poles and zeros. The impedance function has a branch cut \cite{brown2009complex}, because of which the rational approximation has an infinite number of poles and zeros. The branch cut, in this case, can be visualized effectively in the phase portrait of the rational approximation in Fig. \ref{fig:branch_cut}, where one can see a seemingly infinite arc of poles between the branch points \cite{brown2009complex} at zero and infinity.

The distribution of the poles and zeros obtained by the rational approximation of impedance data can thus enable us to understand key properties of the underlying systems. In the next section, we show how rational approximation enables us to derive key properties characteristic of systems with only capacitors and resistors.

\section{Identifiability and uniqueness of equivalent circuits}
\label{sec:identifiability}

In this section, we use rational function approximation to find out if it is possible to recover a circuit from the impedance data it generates. This raises the question: Is there a unique circuit that can produce the impedance dataset, or can multiple circuits produce the same dataset? We focus on circuits with only resistors and capacitors since they are a subset of general circuits in most practical situations. (If it is not possible to recover these simple circuits, adding more complex constant phase elements will not help with unique recovery).

To answer these questions, we first created a library of all the possible circuits made of 5 elements or less, where each element could either be a resistor or a capacitor.  This dataset was created by constructing all possible series-parallel networks with $n$ labeled edges \cite{oeisA006351} where $n \leq 5$ and then reducing them down to unique occurrences among the permutations of the resistors and the permutations of the capacitors. The formulae for the impedance and the poles/zeros of these circuits can be calculated symbolically. This allows us to sort all the circuits by the number of poles/zeros they have. The library of circuits and notebooks showing how to search within them has been made available on GitHub (\url{https://github.com/Dirivian/Rational_EIS}).

\begin{figure}[h]
    \centering
\includegraphics[width=\textwidth,trim={0cm 0cm 0cm 0cm},clip]{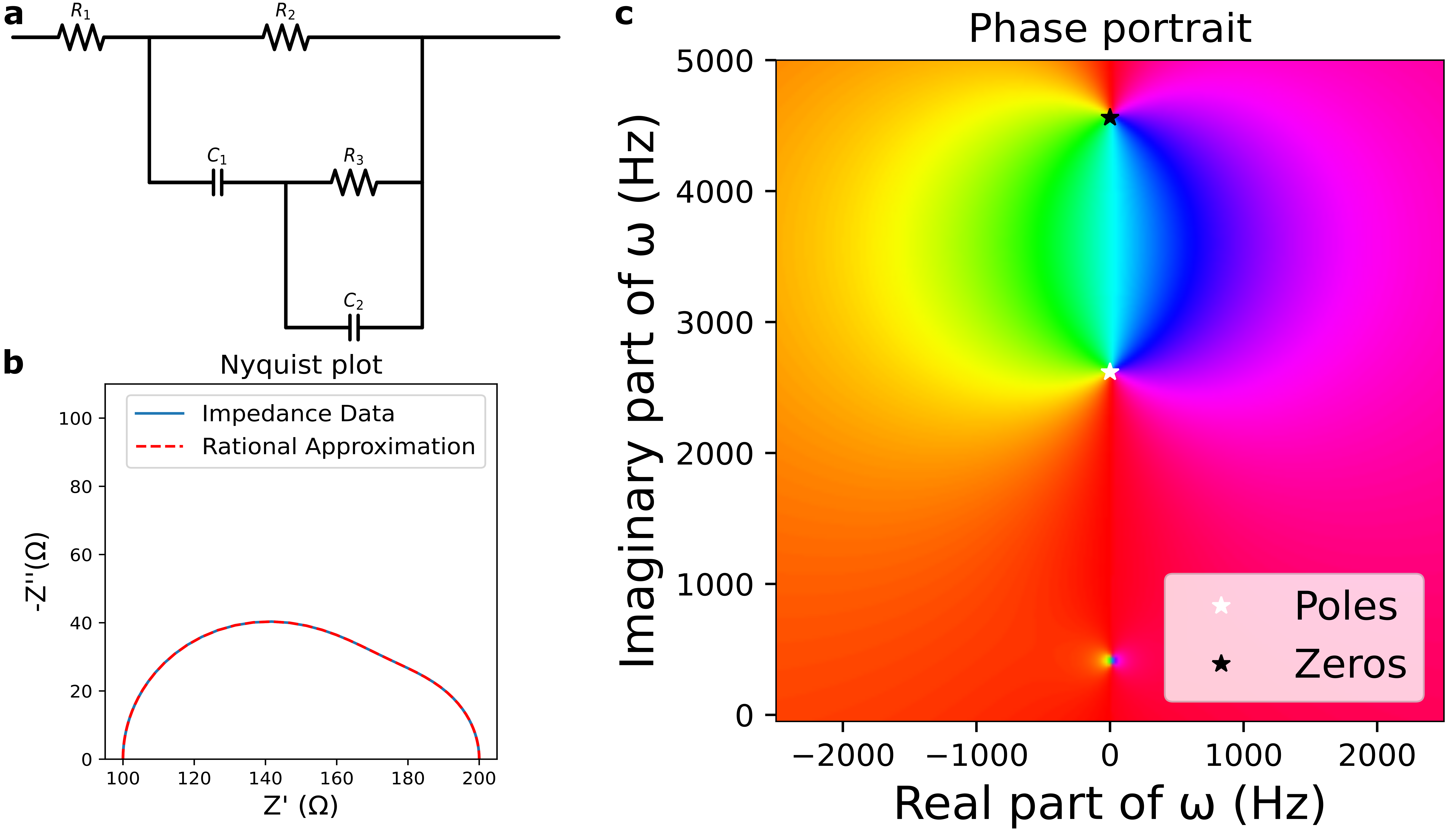}
    \caption{a) A circuit model b) Nyquist plot of the impedance data produced by the circuit model and the rational approximation to it. c) Phase portrait of the rational approximation to the impedance data.}
    \label{fig:circuit1}
\end{figure}

We attempt to recover the circuit from the impedance data generated by the circuit shown in Figure \ref{fig:circuit1}a. The impedance $Z$ depends on the frequency $\omega$ according to Equation \ref{eq:sim_circuits1}. This dataset is generated using 101 values of $\omega$ logarithmically distributed between $1$Hz and $10^6$ Hz. The real and imaginary components of the impedance $Z(\omega)$ can be displayed through a Nyquist plot (Fig \ref{fig:circuit1} b).

\begin{align}
    Z(\omega) = 100 + \frac{1}{\frac{1}{100}+\frac{1}{\frac{1}{10^{-5} i\omega} +\frac{1}{\frac{1}{100}+ 10^{-5} i\omega}}}
    \label{eq:sim_circuits1}
\end{align}

 We can use the AAA algorithm \cite{nakatsukasa2018aaa} to obtain the rational function approximation of the dataset. We see in Fig \ref{fig:circuit1}b that the rational function approximation appears to visually match the impedance dataset. The rational function approximation consists of two poles and two zeros which can be seen in  Fig \ref{fig:circuit1}c. The computed poles and zeros match the exact poles and zeros of Equation \ref{eq:sim_circuits1} to machine precision. 

\begin{figure}[h]
    \centering
    \includegraphics[width=\textwidth]{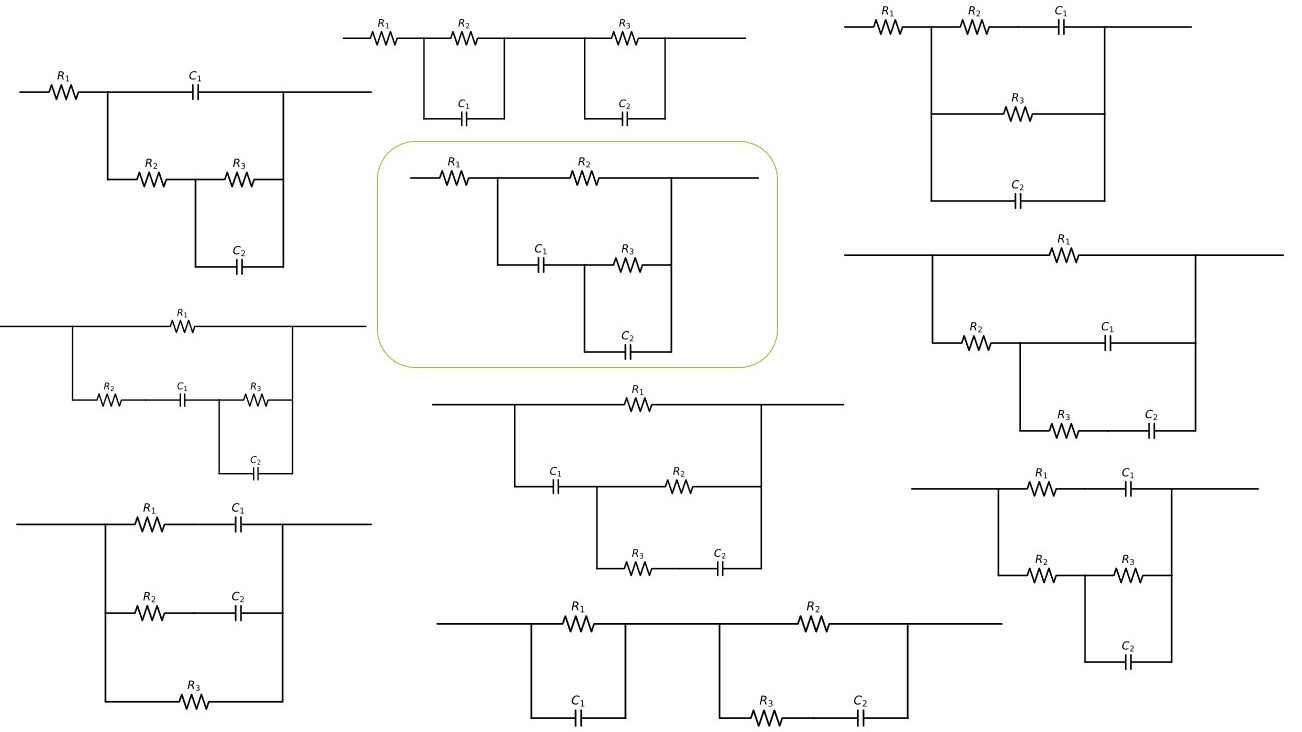}
    \caption{The ten circuits that exactly recreate the impedance data in Fig \ref{fig:circuit1}b. The original circuit (Fig \ref{fig:circuit1}a) is highlighted in the green box.}
    \label{fig:alt_circuits}
\end{figure}

Of the 440 possible circuits in our library, only 48 circuits have two zeros and two poles. Since the computed poles and zeros of the impedance dataset are non-zero, circuits where a pole or zero is symbolically calculated to be zero, can be removed from consideration. This leaves us with ten possible circuits which are shown in Fig \ref{fig:alt_circuits}. All ten circuits have three resistors and two capacitors. Moreover, all their poles and zeros are purely imaginary, just like those of the impedance dataset.

The impedance function of each of these ten circuits has five free variables (three resistances and two capacitances). With five free variables, it is entirely possible to match the four purely imaginary poles and zeros of the impedance data (See Equation \ref{eq:non-unique}). This implies that there are likely multiple circuits that can produce the same impedance data and that the impedance spectrum is not unique.

\begin{align}
Z(\omega; R_1, R_2, R_3, C_1, C_2) =  \frac{K(\omega - z_1 i)(\omega - z_2 i)}{(\omega - p_1 i)(\omega - p_2 i)}
    \label{eq:non-unique}
\end{align}

In this particular case, all ten circuits (Fig \ref{fig:alt_circuits})  can exactly recreate the impedance spectrum shown in Fig \ref{fig:circuit1}.

\section{Properties of circuits with only resistors and capacitors}
\label{sec:properties}

In the previous section, we noted that all the poles and zeros of the circuit in Equation \ref{eq:sim_circuits1} lie on the positive imaginary axis. In this section, we show that circuits consisting only of resistors and capacitors are characterized by poles and zeros that are purely imaginary and non-negative.

The impedance function for a circuit with only resistors, capacitors and inductors can be represented in the general rational form shown in Equation \ref{eq:rational_form}.

\begin{align}
    Z(\omega) = \frac{(i\omega)^n +a_{n-1}(i\omega)^{n-1}+\hdots a_0}{(i\omega)^n +b_{n-1}(i\omega)^{n-1}+\hdots b_0} 
\label{eq:rational_form}
\end{align}

We show in Appendix \ref{si:rc} that Equation \ref{eq:kurtz}/ Equation \ref{eq:kurtz2} are the conditions for all the zeros/poles of the impedance function to be purely imaginary.

\begin{align}
 a_i^2 - 4 a_{i+1}a_{i-1} \geq 0, \quad 1\leq i\leq n-1
    \label{eq:kurtz}
\end{align}

\begin{align}
 b_i^2 - 4 b_{i+1}b_{i-1} \geq 0, \quad 1\leq i\leq n-1
    \label{eq:kurtz2}
\end{align}

If the poles and zeros are purely imaginary, Descartes' rule of signs dictates they must be non-positive. This implies that  if Equation \ref{eq:kurtz}/ Equation \ref{eq:kurtz2} is satisfied, the zeros/poles must lie on the non-negative imaginary axis.

\begin{figure}[h]
    \centering
\includegraphics[width=\textwidth]{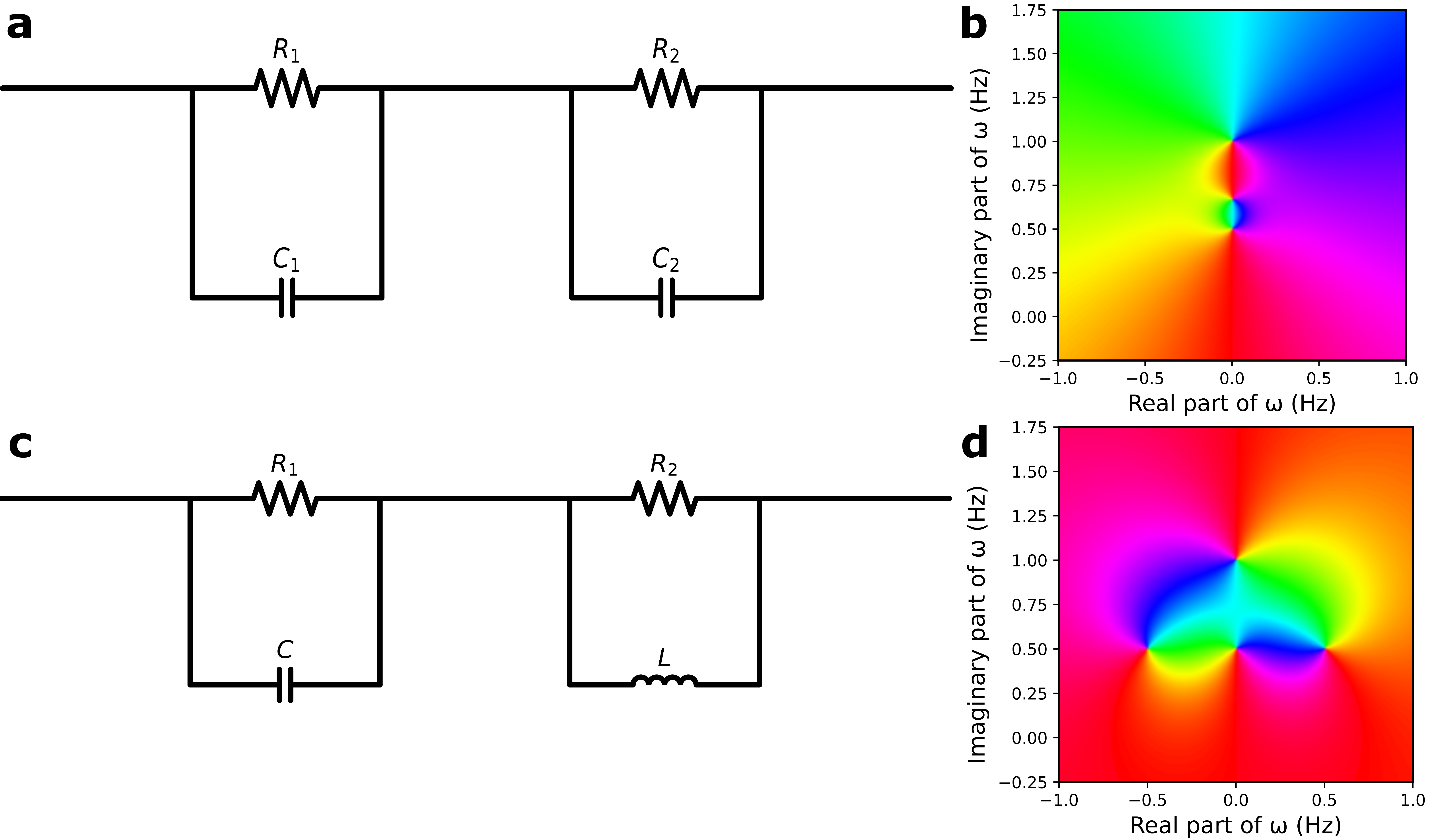}
    \caption{Two similar circuits (a) and (c). The circuit (a) only has resistors or capacitors. Hence, our theory guarantees that its impedance has poles and zeros that lie only on the non-negative imaginary axis (b). Replacing a capacitor in circuit (a) with an  inductor in the bottom circuit (c) can result in  poles and zeros that do not lie on the non-negative imaginary axis (d).}
    \label{fig:capacitor_v_inductor}
\end{figure}

 Consider the top circuit  in Figure \ref{fig:capacitor_v_inductor}. The impedance of this circuit is given by 
\begin{align}
\begin{split}
Z(\omega) = \frac{1}{\frac{1}{R_1}+ C_1 i\omega}    +\frac{1}{\frac{1}{R_2}+ C_2 i\omega} = \frac{(C_1+C_2) i\omega + \frac{1}{R_1} + \frac{1}{R_2}}{ C_1 C_2 (i\omega)^2 +(i\omega)(\frac{C_1}{R_2}+\frac{C_2}{R_1})+\frac{1}{R_1R_2}}   
\end{split}
\end{align}

The numerator has a single root and it lies on the non-negative imaginary axis. For the denominator, we use the coefficients to calculate the left-hand side from Equation \ref{eq:kurtz2}  below and use that to show that poles lie on the non-negative imaginary axis as well.

\begin{align}
\begin{split}
   L.H.S = \bigg(\frac{C_1}{R_2}+\frac{C_2}{R_1} \bigg)^2- 4\frac{C_1C_2}{R_1R_2} =    \bigg(\frac{C_1}{R_2}\bigg)^2+\bigg(\frac{C_2}{R_1} \bigg)^2- 2\frac{C_1C_2}{R_1R_2} 
   = \bigg(\frac{C_1}{R_2}-\frac{C_2}{R_1} \bigg)^2 \geq 0  
\end{split}
\end{align}

For all the circuits from our previously constructed library with 5 elements or less, consisting only of resistors and capacitors, we  can similarly show that equations \ref{eq:kurtz} and \ref{eq:kurtz2} hold. Circuits with only resistors and capacitors thus appear to be defined by the fact that their poles and zeros lie exclusively on the non-negative imaginary axis. While  a general proof for arbitrarily large R-C circuits remains open, no counterexamples have been found.

To highlight that this is not a property of general circuits, consider the circuit on the bottom in Figure \ref{fig:capacitor_v_inductor}. The only difference from the top circuit is that a capacitor is replaced by an inductor. Its impedance is given by
   
\begin{align}
\begin{split}
    Z(\omega) = \frac{1}{\frac{1}{R_1}+ C i\omega}    +\frac{1}{\frac{1}{R_2}+ \frac{1}{ Li\omega}} = \frac{R_1R_2 CL (i\omega)^2 +(i\omega)(R_1+R_2)L+R_1R_2}{ R_1CL (i\omega)^2 +(i\omega)(R_1R_2C+L)+R_2}.
\end{split}
\end{align}

Using the coefficients of the numerator in Equation \ref{eq:kurtz}, we get the following condition for the zeros to lie on the non-negative imaginary axis.
\begin{align}
(R_1+R_2)^2L^2 - 4R_1^2R_2^2CL\geq 0
\end{align}

This condition does not always hold, particularly for large values of $C$. This indicates that circuits with elements that are not purely resistive or capacitive may not satisfy equations \ref{eq:kurtz} and \ref{eq:kurtz2}.

Zeros and poles that do not lie on the non-negative imaginary axis as seen in Figure \ref{fig:capacitor_v_inductor} can thus be evidence of the presence of inductance. Chemical inductors \cite{bisquert2022chemical} which naturally arise in biological and material systems, are often recognized through the phenomenon of `negative capacitance', where dominant inductive behavior causes impedance data to appear in the fourth quadrant of the Nyquist plot. Though our primary aim in this section was to explore the properties of circuits with only resistors and capacitors, the pole-zero perspective developed here  provides a complementary direction for identifying chemical inductors, particularly in systems where the capacitive elements are known to be ideal.

\section{Detecting  and identifying Constant Phase Elements}
\label{sec:CPE}

\begin{figure}
    \centering
    \includegraphics[width=\textwidth]{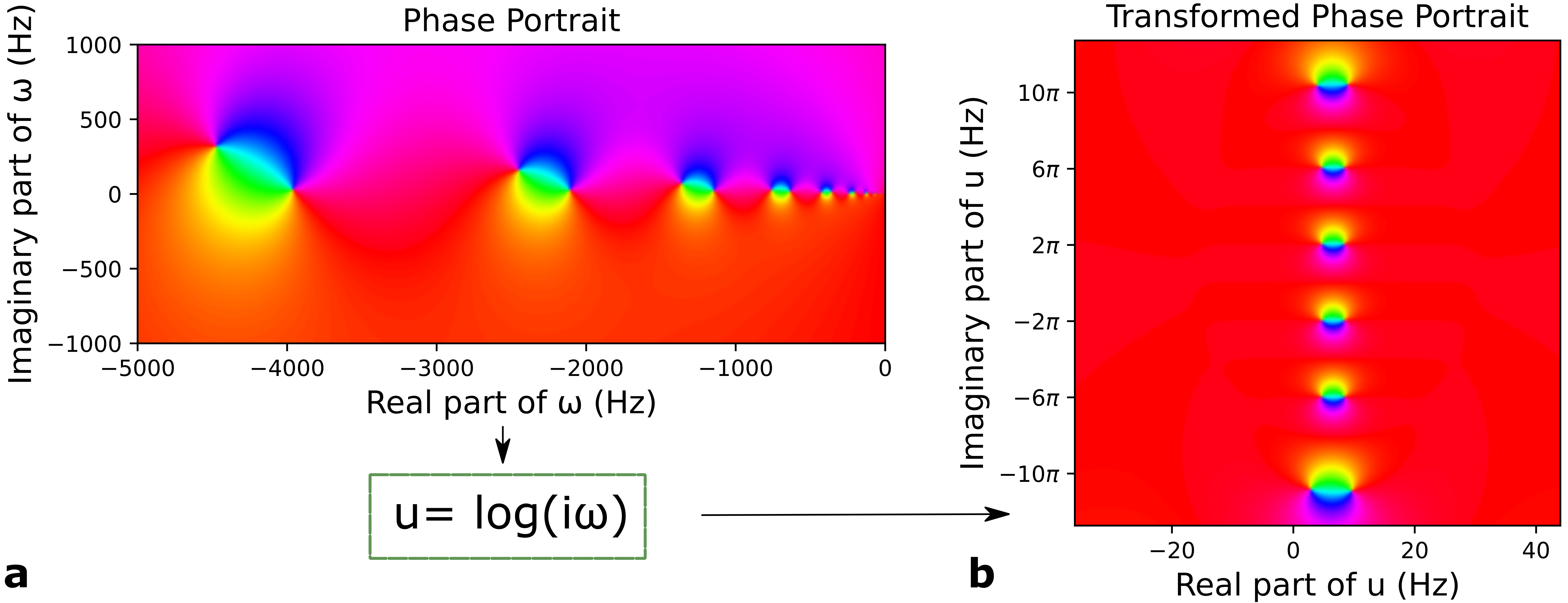}
    \caption{ Circuits with constant phase elements are visually characterized by seemingly infinite series of poles and zeros. a) Phase portrait of the impedance given by Equation \ref{eq:simple_CPE2}. The log substitution allows us to obtain the transformed phase portrait (b).  The presence of poles and zeros with imaginary components at $2(\pi i +2n\pi i)$ allow us to deduce that $\alpha$ in Equation \ref{eq:simple_CPE2} is $\frac{1}{2}$.}
    \label{fig:detect_identify_CPE}
\end{figure}

The presence of constant phase elements almost always results in poles and zeros that do not lie on the non-negative imaginary axis. Typically, it is possible to visually identify constant phase elements by looking for branch cuts in the phase portraits of the rational approximation (See Fig. \ref{fig:branch_cut}). Consider a simple circuit with a constant phase element with impedance given by 

\begin{align}
    Z(\omega) = R_1 + \frac{1}{\frac{1}{R_2}+ A (i \omega)^\alpha}
    \label{eq:simple_CPE2}
\end{align}

The branch cut, visualized by the series of poles and zeros in Fig. \ref{fig:detect_identify_CPE}, allows us to infer the presence of a constant phase element.
Once a constant phase element is detected in this manner, it is also possible to identify its phase. We employ the substitution $u = \log(i\omega)$ to obtain a transformed impedance function (Eq. \ref{eq:simple_CPE3}).

\begin{align}
\begin{split}
    Z(u) = R_1 + \frac{1}{\frac{1}{R_2}+ A \exp(\alpha u)} = \frac{\frac{R_1}{R_2}+ A R_1 \exp(\alpha u)+ 1}{\frac{1}{R_2}+ A \exp(\alpha u)}
\end{split}
     \label{eq:simple_CPE3}
\end{align}

The poles and zeros of this transformed impedance function lie at $\frac{1}{\alpha}(\pi i +2n\pi i)$ along the imaginary direction (Figure \ref{fig:detect_identify_CPE}) for different integer values of $n$ (See Appendix \ref{appendix:transformed_CPE}). This means that the location of the poles and zeros of the transformed impedance function can be used to identify $\alpha$, the phase of the constant phase element. (See Figure \ref{fig:detect_identify_CPE}). Note that while theoretically, there must be an infinite number of poles and zeros at  $\frac{1}{\alpha}(\pi i +2n\pi i)$ for all integer values of $n$, there is only a finite amount of impedance data which is further complicated by measurement noise.  So, in practice, we typically only observe the poles and zeros close to the real line ($-1 \leq n \leq 0$).

 The log substitution approach allows us to identify the constant phase element in Equation \ref{eq:simple_CPE2}. In more complex circuits with multiple capacitive and constant phase elements, overlapping features makes identification harder. In such cases, isolating frequency regimes where one element dominates becomes necessary. We discuss these scenarios and associated strategies in detail in Appendix \ref{appendix:secCPE_multiple}.

\section{Application to impedance response measurements from a lithium-ion battery  coin cell }
\label{sec:applications}

\begin{figure}
    \centering
    \includegraphics[width=\textwidth]{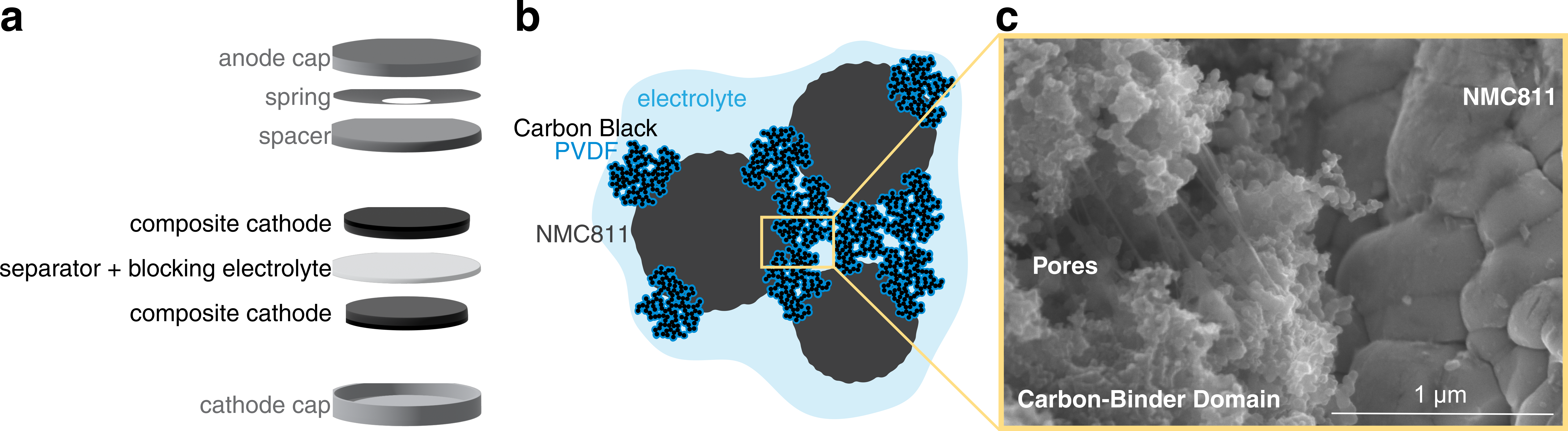}
\caption{a) Symmetric (two composite cathodes) coin cell stack under blocking electrolyte conditions b) Depiction of the composite cathode microstructure including NMC811, Carbon Black/PVDF domain (Carbon-Binder Domain), and pore space wet by the electrolyte. c) Scanning Electron Microscopy (SEM) micrograph of the composite cathode microstructure.}
    \label{fig:lopez_example}
\end{figure}

In this section, we apply the  mathematical framework presented in this paper to real-world data. We look at  impedance response measurements from a lithium-ion battery  coin cell \cite{liu2024structure} to understand the electronic properties and structure of a lithium-ion battery  composite cathode (Fig. \ref{fig:lopez_example}). 

The cathode is a porous film consisting of a lithium metal oxide (LiNi$_{0.8}$Mn$_{0.1}$Co$_{0.1}$O$_2$), carbon black, and polymer binder (polyvinylidene fluoride (PVDF)). During traditional lithium-ion battery  operation, a lithium-salt electrolyte wets the cathode pores to transport positively charged Li ions to the surface of metal oxide particles.  Li ions then reversibly intercalate into the layered structure during discharge and charge. This process is balanced by the transport of negatively charged electrons through the carbon black electronic network. The impedance measurements are often used to understand the cathode capacitance, charge-transfer resistance, and ionic resistance through the pores \cite{aada5d06d3e74a70b243e3a0fa7f98c9, article}.

In this particular setup, a  symmetric coin cell is used with the porous cathode as the positive and negative electrode, in order to isolate the impedance response of the cathode. A Li-free blocking electrolyte (10 mmol tetra-butyl ammonium hexafluorophosphate (TBAPF6) in ethylene carbonate (EC) and dimethyl carbonate (DMC) in a 1:1 weight ratio) is used, preventing intercalation into the metal oxide. This allows us to simplify the solid electrode - electrolyte interfacial impedance behavior \cite{article}.

\begin{figure}[ht!]
    \centering
    \includegraphics[width=\textwidth]{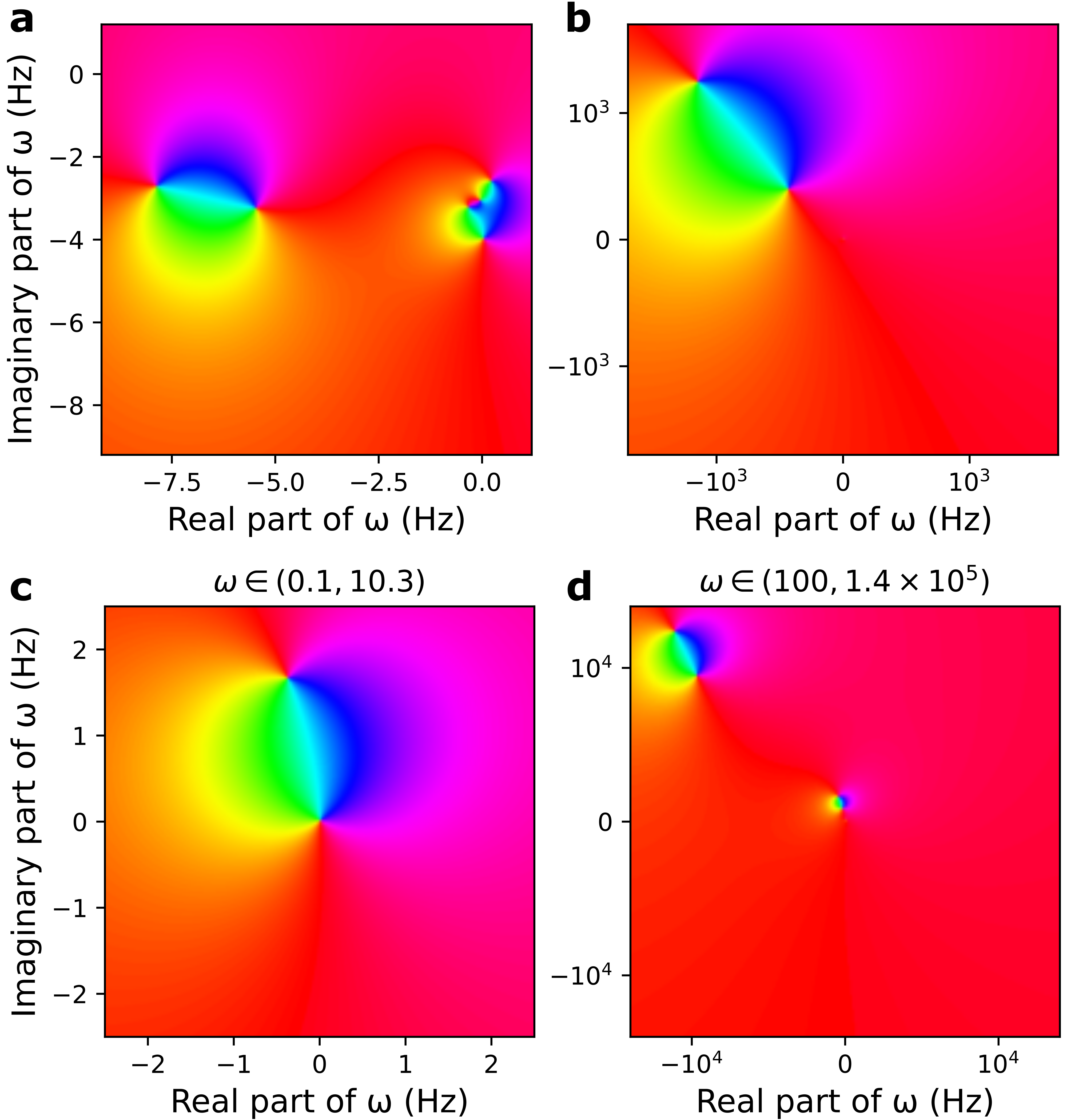}
\caption{ a) Phase portrait of the rational approximation to impedance data from a lithium-ion battery  coin cell. We see complicated pole-structure across different scales. b) Phase portraits after separating the impedance data into low and high frequency regimes. While the high frequency regime still shows a variety of features at different scales, the low-frequency regime has a simple structure.}
    \label{fig:lopez_total_pp}
\end{figure}

Rational approximation of the full impedance spectrum reveals a complicated pole-zero structure (Figure \ref{fig:lopez_total_pp}a and b), with poles and zeros scattered across the complex plane in wildly different regions. This is not surprising, since diffusion  of ions through a porous electrode is a non-trivial process, mathematically modeled by the transmission line model \cite{lazanas2023electrochemical} given by Equation \ref{eq:transmission} where $R_{ion}$ is the ionic resistance and $1/Q(i \omega)^\alpha$ is the impedance corresponding to the distributed capacitance. 
\begin{align}
Z_{TLM}(\omega) = \sqrt{\frac{R_{ion}}{Q(i\omega)^\alpha}}\coth(\sqrt{R_{ion}Q(i\omega)^\alpha}) 
    \label{eq:transmission}
\end{align}

The square-root and the hyperbolic cotangent functions in the transmission line model results in a complicated pole-zero structure, which explains Figure \ref{fig:lopez_total_pp}a and b. 

The impedance in our system can be represented  by Equation \ref{eq:transmission+M}, where the first term is the transmission line model and the second term $M(\omega)$ represents the remaining impedance effects in the system.

\begin{align}
Z(\omega) = Z_{TLM}(\omega) + M(\omega)
    \label{eq:transmission+M}
\end{align}

Since $R_{ion}, Q, \alpha$ and $M(\omega)$ are unknown, Equation \ref{eq:transmission+M} is an incomplete mathematical description. The standard EIS approach to characterize such a system involves making assumptions about $R_{ion}, Q, \alpha$ and $M(\omega)$. Each assumption opens up an additional dimension of uncertainty. With these assumptions and enough noise, searches for models in a highly non-convex optimization landscape can show good fits for non-representative models. 

Our goal is to use information in the impedance measurements from different frequency regimes (Figure \ref{fig:lopez_total_pp}c and d) to minimize the uncertainty in the search of models that characterize our material systems. In the low-frequency regime, the rational approximation visualized in Figure \ref{fig:lopez_total_pp}c  reveals a much simpler structure, a pole and a zero  lying very close to the positive imaginary axis, indicating a circuit model given by Equation \ref{eq:R_C_low}. This corresponds to a resistor with resistance $R$ and an imperfect capacitor with impedance $Z(\omega) = 1/C(i \omega)^\alpha$ in series. This low-frequency model allows us to find the $Q$ and $\alpha$ values in Equation  \ref{eq:transmission} (See Appendix \ref{appendix:data} for details).

\begin{align}
Z(\omega) =  R + \frac{1}{C (i \omega)^\alpha}  
    \label{eq:R_C_low}
\end{align}

\begin{figure}[ht!]
    \centering
    \includegraphics[width=0.95\textwidth]{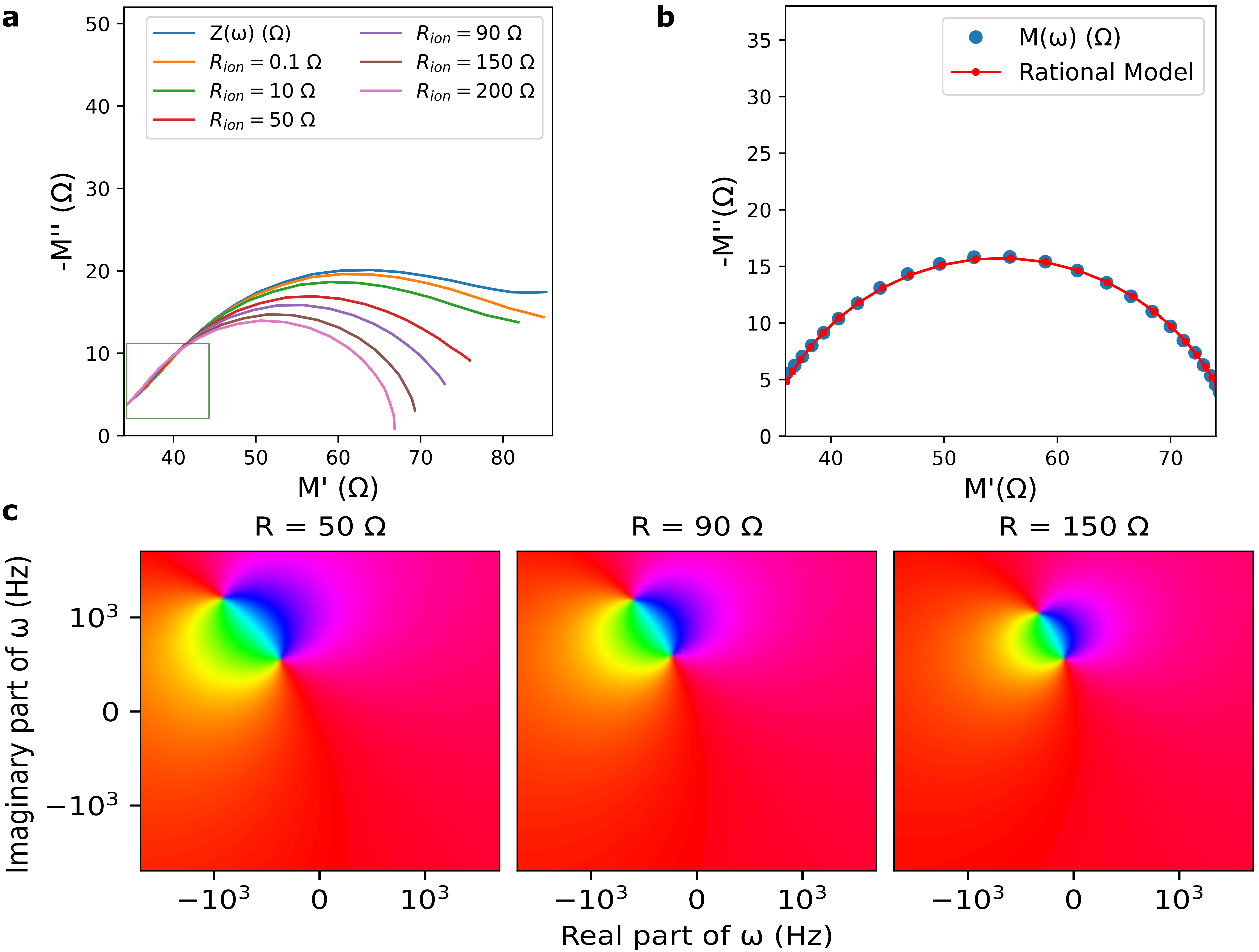}
\caption{ a) Nyquist plots of $M(\omega)$ (Equation \ref{eq:M}) for different $R_{ion}$ choices. The blue curve is the original impedance data $Z(\omega)$. The green rectangle is used to highlight the frequency regime which appears independent of $R_{ion}$. b) A simple  model provides a good rational approximation to $M(\omega)$ (for $R_{ion} = 90 \Omega$).  c) Phase portraits of rational approximations to $M(\omega)$ for different $R_{ion}$ choices highlight a similar pole-zero structure despite large variation in $R_{ion}$ values. }
    \label{fig:M_plots}
\end{figure}

$R_{ion}$ cannot be similarly identified because $M(\omega)$ likely has resistive contributions in the low-frequency regime. This reveals a possible degree of mathematical uncertainty in the estimation of $R_{ion}$. To explore this, we can isolate $M(\omega)$ via

\begin{align}
    M(\omega) = Z(\omega) -Z_{TLM}(\omega, R_{ion})
    \label{eq:M}
\end{align}
 and analyze its  Nyquist plots and phase portraits for different values of $R_{ion}$ (Figure \ref{fig:M_plots}).

We see in Figure \ref{fig:M_plots} that while the Nyquist plots of $M(\omega)$ vary noticeably for different $R_{ion}$  values, the features in the phase portraits remain roughly the same. This illustrates that moderate differences in the $R_{ion}$ values  do not lead  to significant changes in the underlying model. We also notice, in Figure \ref{fig:M_plots}a, that there is a frequency regime (highlighted by the green rectangle) where the impedance does not appear to depend on the choice of $R_{ion}$. Sampling this regime allows reliable modeling of $M(\omega)$ but provides little information about $R_{ion}$ itself.

For various choices of $R_{ion}$ near 90 Ohms, we obtain a simple rational model of the form given by Equation \ref{eq:M_model}

\begin{align}
    M(\omega) = \frac{a(i \omega)^{\beta}+b}{(i \omega)^{\beta}+c}
    \label{eq:M_model}
\end{align}

 Figure \ref{fig:M_plots}b shows that this model  matches the $M(\omega)$ data very well for a representative choice of $R_{ion}= 90 \Omega$ ($a =34, b = 16496, c =218, \beta = 0.82$). With $M(\omega)$  recovered, we have Equation \ref{eq:complete_model}, a complete mathematical description of the impedance response from the lithium-ion battery  coin cell.

 \begin{align}
\begin{split}
    Z(\omega) = \frac{a(i \omega)^{\beta}+b}{(i \omega)^{\beta}+c} +  \sqrt{\frac{R_{ion}}{Q(i\omega)^\alpha}}\coth(\sqrt{R_{ion}Q(i\omega)^\alpha}) 
\end{split}
    \label{eq:complete_model}
\end{align}

\begin{figure}[ht!]
    \centering
\includegraphics[width=\textwidth]{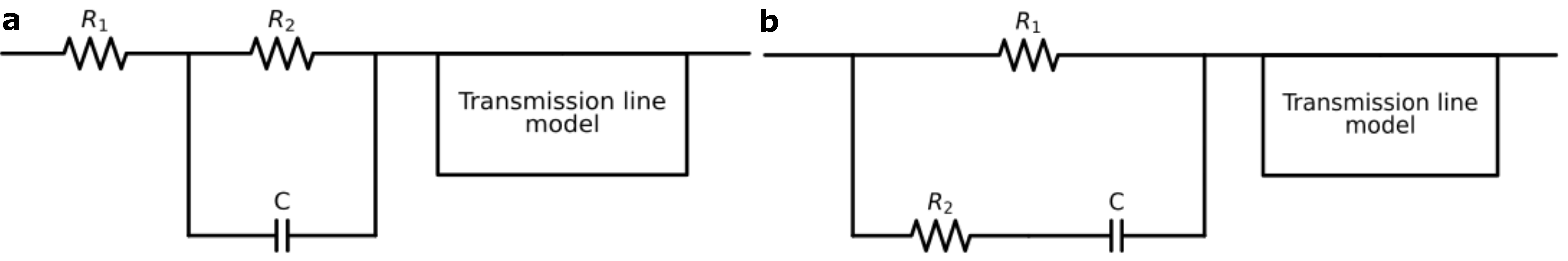}
\caption{The two simplest circuit models that explain the impedance data.}
    \label{fig:final_model_circuits}
\end{figure}

 There are two minimal circuits (Figure \ref{fig:final_model_circuits}) that share the same pole structure  and thus result in the same mathematical model, Equation \ref{eq:complete_model}.

\begin{figure}[ht!]
    \centering
\includegraphics[width=\textwidth]{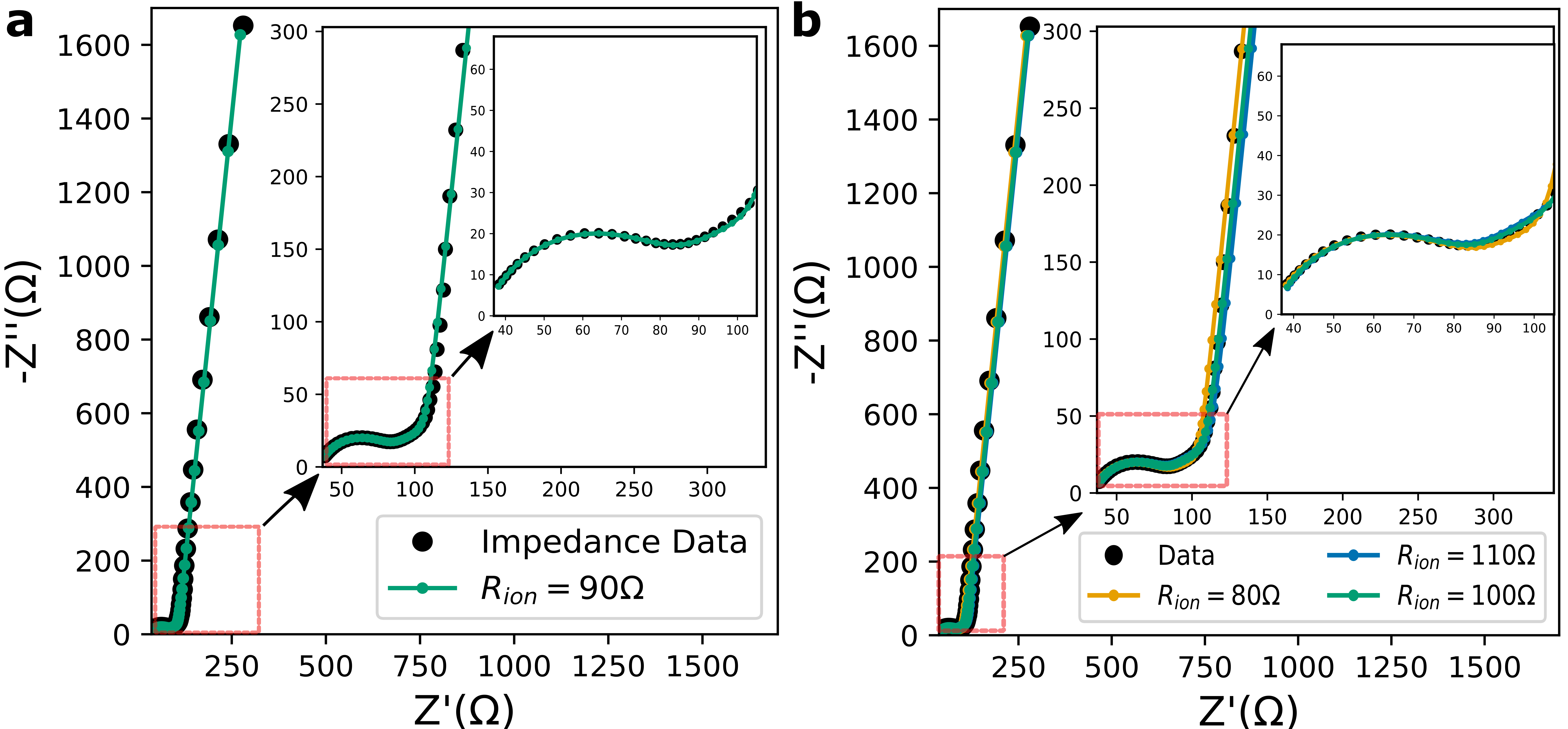}
\caption{a) The impedance data is compared with the derived mathematical model (Equation \ref{eq:complete_model}) b) $R_{ion}$ is varied and the ensuing mathematical models are compared. This reveals a degree of mathematical uncertainty inherent in the estimation of $R_{ion}$.}
    \label{fig:final_model}
\end{figure}

In Figure \ref{fig:final_model}a and b, we see that this derived mathematical model agrees very well with the impedance measurements for two different choices of $R_{ion} = 90 \Omega, 100 \Omega$. Note that no conventional curve fitting has been done in Figure \ref{fig:final_model}. Instead,  the parameters for the models are directly obtained from the rational approximation (See Appendix \ref{appendix:parameters} for the parameter values). 

Only when $R_{ion}$ is pushed beyond 80 and 110 Ohms, do we start to see some deviation of the derived models from the impedance measurements. This indicates that there exists a continuous family of models, parameterized by values of $R_{ion}$ near 90 Ohms that match the impedance measurements very well. Our analysis thus reveals that the process of recovering $R_{ion}$ through impedance analysis has a degree of inherent mathematical uncertainty associated with it. 

Furthermore, as we see in Fig. \ref{fig:final_model}b, the deviations of the model from the measurements when changing $R_{ion}$ are most pronounced in the intermediate frequency range (1 to 100 Hz). It is very important that the measurements in that frequency range be precisely collected, so as to avoid compounding the intrinsic mathematical uncertainty with measurement error.

This example demonstrates the strength of a model-agnostic, data-driven approach to impedance analysis. By extracting features directly from the data, without relying on predefined circuit templates, we were able to identify two minimal circuit models (Fig. \ref{fig:final_model_circuits}) that accurately reproduce the measured impedance response. This same approach also revealed that a range of $R_{ion}$ values can lead to equally valid models. 

In the context of lithium-ion battery systems, $R_{ion}$ is used to calculate electrode tortuosity, a key determinant of ionic transport within the porous structure \cite{article}. While our results show that moderate variations in $R_{ion}$ do not strongly affect the impedance fit, they do impact the inferred tortuosity values, which are directly proportional to $R_{ion}$. This mathematical uncertainty introduces challenges in repeatability of tortuosity determination across different fitting methods. We conclude this uncertainty should be explicitly reported and propagated into final tortuosity estimates, and the details of the fitting algorithm used to extract $R_{ion}$ should be clearly documented.

\section{Discussion and Outlook}

This paper introduces a mathematical framework based on rational approximation to understand and interpret EIS data. A central highlight of this framework is that it enables us to extract key characteristic features, poles and zeros, directly from the data without assuming a model. In contrast, traditional EIS analysis typically begins with a predefined model structure. We have shown throughout this paper, that it is these features that uniquely characterize the data and not the fitted circuits which may be non-unique. Once these features are extracted from the data, we can then search for models that share the same features, enabling a more targeted and informed model selection process.

Alongside this data-driven strategy, this paper also introduces mathematical theory to distinguish between different classes of models. This enables us to distinguish circuit models with only resistors and capacitors from models with inductors and constant phase elements. Chemical inductors are typically identified by the presence of `negative capacitance' \cite{lazanas2023electrochemical, bisquert2022chemical}. Our analysis shows promise for detecting inductive behavior even in scenarios when it is not dominant enough to produce `negative capacitance'. 

The theory developed in this paper to characterize different classes of models can also be used to identify model transitions in complex material systems. Identifying such transition points can reveal key physical properties of the underlying system. For example, in thin-film diffusion impedance \cite{doi:10.1021/acs.jpcc.1c04572}, the turnover frequency provides insight into the system’s natural frequency and diffusion coefficient. Using the tools developed here, this turnover point can be identified as the frequency at which the pole-zero structure shifts from that  of a simple R-C circuit to one exhibiting the branch-cut behavior associated with a constant phase element.

For the case of circuits with resistors and capacitors, we show that the poles and zeros lie on the non-negative imaginary axis. This constraint reduces the pole-zero structure to a one-dimensional axis, making it possible to enumerate all possible models (up to a certain number of circuit elements) using just the number of poles and zeros. We can thus isolate parts of the impedance dataset where the impedance is largely resistive and capacitive and search in our library of circuits for the family of models that satisfy the  observed pole-zero structure.

For circuits with inductors and constant phase elements, there are two major issues with enumerating all possible circuits. First of all, with finite data, it is  not always possible to distinguish inductive elements from constant phase elements. A constant phase element is typically characterized by a branch cut, which is an arc of theoretically infinite poles and zeros in the complex plane. However, since impedance data is finite and complicated by noise, the number of poles and zeros are finite and sometimes, may be too few to distinguish them from the poles and zeros of inductors. In the future, we envision developing additional mathematical theory for distinguishing the features of inductive and constant phase elements.

Secondly, the inclusion of constant phase elements dramatically increases the size of the space of all possible circuits since  the phase of a constant phase element can assume a continuum of values.  For simple circuits, the log substitution approach presented in Section \ref{sec:CPE} allows us to identify the phase of the constant phase elements. However, in a circuit model with multiple contributing elements, it is difficult to de-convolve the individual contributions of each element (See Appendix \ref{appendix:secCPE_multiple}). In this paper, we have employed the idea that different components are dominant in different frequency regimes to isolate and identify them. This, in turn, informs us of which frequency regimes to sample to get the necessary information about each component. This perspective also highlights the importance of strategic frequency sampling: by targeting the regimes where specific components are most influential, we can maximize the information content of the measurements. While much attention in the EIS community has focused on fitting models to existing data, relatively little work has addressed the complementary problem of identifying the most informative regions to sample. The analysis presented here may serve as a foundation for advancing this area.

One important consideration in our current framework is that the rational function approximation does not explicitly account for noise in the data. Since noise is inevitable in most experimental datasets, a rational function approximation with too many poles and zeros can end up overfitting the noise and resulting in erroneous features. With too few poles and zeros, key features of the datasets may be missed. As a result, a degree of manual fine-tuning is currently required to capture the essential structure of the dataset without overfitting. Because of this, our analysis pipeline is not yet fully automated and still relies on human input and judgment. There has been substantial work in the EIS community on using Kramers-Kronig (KK) relations \cite{boukamp1995linear, urquidi1990applications} to assess and quantify noise levels in impedance data. Incorporating uncertainty information from KK tests into the rational approximation process, in combination with robust model selection techniques, represents a promising direction for developing a fully automated and noise-aware implementation of this framework.

Finally, the techniques presented in this paper are designed to complement, rather than replace, standard EIS methods. This paper has largely focused on circuit models and has highlighted new avenues for model discovery and interpretation through data-driven feature-based exploration of broad families of models. Looking ahead, integrating this framework with insights from 
Distribution of Relaxation Times (DRT) methods \cite{ciucci2019modeling, boukamp2018use, dierickx2020distribution, lu2022timescale} offers a promising path toward a more comprehensive and refined interpretation of EIS spectra. 

\section{Conclusion}

In this work, we demonstrated how numerical rational function approximation provides a powerful framework for interpreting impedance measurements. Our main findings can be summarized as follows:

\begin{itemize}
    \item Poles and zeros can be extracted directly from the data and serve as unique signatures of the impedance spectra. While different equivalent circuit models may reproduce the same spectra, the poles and zeros must coincide across such models. 
    \item Constant phase elements can often be identified through the appearance of branch cuts in phase portraits. We also developed mathematical approaches to further characterize such behavior, which work reliably for simple circuits but require careful analysis and additional information for more complex systems.
    \item For purely resistive–capacitive systems, poles and zeros lie on the non-negative imaginary axis. This property enables us to isolate systems or frequency regimes where the impedance is dominated by resistive and capacitive contributions.
    \item In some cases, the approximation reveals poles and zeros with significant real parts, even in the absence of a branch cut. Such behavior can often point to the presence of a chemical inductor. 
    \item Application of the framework to an experimental lithium-ion battery dataset provided direct insights into the system, particularly about the uncertainty associated with estimating $R_{ion}$.
    \item Finally, we highlighted the limitations of the present approach and outlined directions for future work, including algorithmic improvements and applications for distinguishing between competing models. 
\end{itemize}
In conclusion, this work presents a new way to think about how we extract and interpret information from impedance data and we hope it inspires novel pathways for understanding material systems.
\section*{Acknowledgments}
The authors would like to thank Prof. Heather Wilber for valuable discussions on algorithms for numerical extraction of poles and zeros and for directing us to the AAA algorithm and Prony's method. This work is primarily supported by the National Science Foundation Materials Research Science and Engineering Center at Northwestern University (Award Number: NSF DMR-2308691). J.L. acknowledges support from the U.S. Department of Energy (DOE), Vehicle Technologies Office (VTO) with award number DE-EE0009644. W.B. acknowledges support by the National Science Foundation Graduate Research Fellowship under Grant No. (DGE-2234667). The SEM image in Fig \ref{fig:lopez_example} was made using the EPIC facility (RRID: SCR\_026361) of Northwestern University’s NUANCE Center, which has received support from the SHyNE Resource (NSF ECCS-2025633), the IIN, and Northwestern's MRSEC program (NSF DMR-2308691). Work performed at the Center for Nanoscale Materials, a U.S. Department of Energy Office of Science User Facility, was supported by the U.S. DOE, Office of Basic Energy Sciences, under Contract No. DE-AC02-06CH11357.

\bibliographystyle{siamplain}
\bibliography{references}

 \appendix
 \section{ Properties of circuits with only resistors and capacitors} \label{si:rc}

The impedance function for a circuit with only resistors, capacitors, and inductors can be  
represented in a general rational form as
\begin{align}
    Z(\omega) = \frac{(i\omega)^n +a_{n-1}(i\omega)^{n-1}+\hdots a_0}{(i\omega)^n +b_{n-1}(i\omega)^{n-1}+\hdots b_0} =  \frac{y^n +a_{n-1}y^{n-1}+\hdots a_0}{y^n +b_{n-1}y^{n-1}+\hdots b_0}
\end{align}

The roots of the numerator are the zeros of the impedance function, and the roots of the denominator are the poles of the impedance function. 

Consider the polynomial $y^n +a_{n-1}y^{n-1}+\hdots a_0$ from the numerator of the above equation. Since resistances and capacitances have to be positive, the coefficients $a_n$ and $b_n$ are positive. From \cite{Kurtz01031992},  we know Equation \ref{eq:kurtz_appendix} needs to be satisfied for the roots of a polynomial with positive coefficients $a_n$ to be purely real. 
\begin{align}
 a_i^2 - 4 a_{i+1}a_{i-1} \geq 0, \quad 1\leq i\leq n-1
    \label{eq:kurtz_appendix}
\end{align}

If the roots of the polynomial  $y^n +a_{n-1}y^{n-1}+\hdots a_0$ are all real, since  the coefficients $a_n$ are positive, Descartes rule of signs implies that all the roots are negative (or zero).

Since $y = i\omega$, Equation \ref{eq:kurtz_appendix} is thus the condition for the poles of the impedance function to lie on the non-negative imaginary axis. The zeros can similarly be proven to lie on the non-negative imaginary axis.

\section{Poles and zeros of the transformed impedance function} \label{appendix:transformed_CPE}

Consider the transformed impedance function given in Equation 15 of the main text and written below as Equation \ref{eq:simple_CPE3_appendix}

\begin{align}
    Z(u) = R_1 + \frac{1}{\frac{1}{R_2}+ A \exp(\alpha u)} =\frac{\frac{R_1}{R_2}+ A R_1 \exp(\alpha u)+ 1}{\frac{1}{R_2}+ A \exp(\alpha u)}
     \label{eq:simple_CPE3_appendix}
\end{align}

Let $u^*$ denote any root of the denominator in Equation \ref{eq:simple_CPE3_appendix}.

\begin{align}
  \exp(\alpha u^*) &=  -\frac{1}{AR_2} \label{eq:simple_CPE4}
  \end{align}
  
Since all the coefficients in our circuits are non-negative, Descartes' Rule of Signs implies if $\exp(\alpha u^*)$ is real, then $\exp(\alpha u^*)$ cannot be positive. This results in a negative sign being present in the right-hand side of Equation \ref{eq:simple_CPE4}. When we solve Equation \ref{eq:simple_CPE4} for $u^*$, the negative sign causes a $\pi$ i term to appear.

\begin{align}
u^* &=  \frac{1}{\alpha}\log(\frac{1}{AR_2}) + \frac{1}{\alpha}(\pi i +2n\pi i), \quad n = 0, \pm 1, \pm 2, \hdots
\end{align}

The presence of the $\pi$ i term enables us to extract the $\alpha$ value by looking at the imaginary part of $u^*$, the poles of the transformed impedance function. We can repeat this process to show that the zeros of the transformed impedance function also lie at $\frac{1}{\alpha}(\pi i +2n\pi i)$ in the imaginary direction.

Note that since the log function is multi-valued, there are infinitely many poles. However, rational function approximation typically recovers only a few of the poles and zeros close to the origin since that is sufficient for a good approximation for finite data.

\section{  Prony's method.}\label{appendix:prony}

Prony's method \cite{prony1795essai, weiss1963prony} is a linear algebra technique that allows us to recover the exponents $\alpha_k$ and the coefficients $c_k$ from measurements $f(x)$ of the form

\begin{align}
f(x) = \sum_k c_k \exp(\alpha_k x).
\label{eq:prony_sup}
\end{align}

Given $k+1$ equispaced samples, the structure of Equation \ref{eq:prony_sup} implies that there exist coefficients $p_i$ such that
\begin{align}
    \sum_{i=0}^{k} p_i f(x_0 + i\Delta x) = 0.
    \label{eq:pronyeq1}
\end{align}
Since this relation holds for arbitrary $x_0$, we can collect multiple instances of $f(x_0+i\Delta x)$ for different $x_0$ values and solve a linear system to determine the coefficients $p_i$.

Again, due to  the structure of Equation \ref{eq:prony_sup}, the roots of the polynomial
\begin{align}
   \sum_{i=0}^{k} p_i z^{i} = 0
   \label{eq:pronyeq2}
\end{align}
are  given by $z = \exp(\alpha_k)$. Thus, solving Equation~\ref{eq:pronyeq1} provides the $p_i$, and the roots of Equation~\ref{eq:pronyeq2} yield the $\alpha_k$. Once the $\alpha_k$ are known, Equation~\ref{eq:prony_sup} becomes a linear system in the coefficients $c_k$, which can then be determined.

Let us demonstrate this using a simple algebraic example. Suppose
\begin{align}
    f(x) = e^x + e^{2x}.
\end{align}
To recover this expansion, we first compute the $p_i$. Taking $x_0 = -1,0,1$ and $\Delta x = 1$, we obtain a system of equations corresponding to Equation~\ref{eq:pronyeq1}. In explicit form, this reads
\begin{align}
\begin{bmatrix}
    f(-1) & f(0) & f(1) \\
    f(0) & f(1) & f(2) \\
    f(1) & f(2) & f(3)
\end{bmatrix}
\begin{bmatrix}
    p_0 \\ p_1 \\ p_2
\end{bmatrix}
= 
\begin{bmatrix}
    0 \\ 0 \\ 0
\end{bmatrix}.
\end{align}
The vector $[p_0, p_1, p_2]$ lies in the null space of this matrix, and in practice it can be computed as the right singular vector corresponding to the zero singular value. For this simple case, one can algebraically verify that
\begin{align}
    [p_0, p_1, p_2] = [e^3,\; -(e+e^2),\; 1].
\end{align}

The corresponding polynomial is
\begin{align}
    z^2 - (e+e^2)z + e^3 = 0.
\end{align}
Its roots are $z = e$ and $z = e^2$. Since $z = \exp(\alpha_k)$, we obtain $\alpha_k = 1,2$, which matches the exponents in the original function. With the $\alpha_k$ values known, the coefficients $c_k$ can then be solved for using Equation~\ref{eq:prony_sup}, yielding $c_1=c_2=1$.

In practice, due to imperfect data, Equation~\ref{eq:pronyeq1} may not hold exactly. Thus, the regularized Prony method \cite{wilber2022data} is typically employed, in which Equation~\ref{eq:pronyeq1} is solved for the right singular vector corresponding to the smallest singular value, which may be nonzero but is required to lie below a prescribed threshold. Moreover, the number of exponential terms $k$ is not known in  advance. One typically tries increasing values of $k$ until a nontrivial null vector is found. For instance, in the example above no such null vector exists for $k=1$. If $k > 2$ is assumed, the resulting null vector corresponds to additional zero terms (e.g., $f(x) = e^x + e^{2x} + 0 + 0$). A useful process to verify that  the data indeed has the form of Equation~\ref{eq:prony_sup}. is to check that the nonzero $\alpha_k$ values remain consistent as $k$ is increased.

\section{ Identifying Constant Phase Elements when multiple competing elements are present. }\label{appendix:secCPE_multiple}

 The log substitution approach allows us to identify the constant phase element in Equation 14 of the main text. However, the process is not as straightforward when the circuit has other capacitive and constant phase elements.
 
 Consider the standard Randles circuit \cite{randles1947kinetics} with the impedance given by Equation \ref{eq:randles}.
\begin{align}
Z(\omega) = R_1 + \frac{1}{C i \omega + \frac{1}{R_2 + 1/A\sqrt{i \omega}}}
    \label{eq:randles}
\end{align}

\begin{figure}
    \centering
    \includegraphics{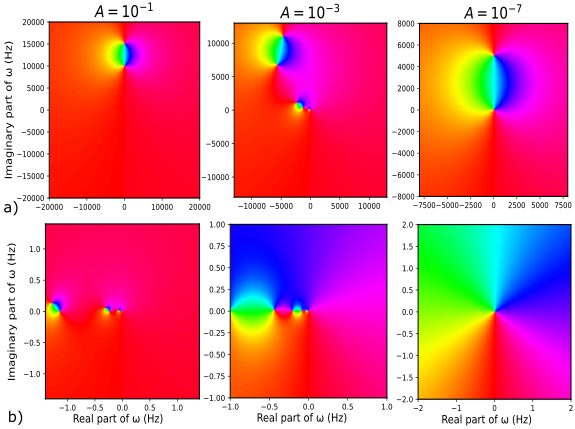}
    \caption{a) Phase portraits of the Randles circuit (Equation \ref{eq:randles}, $ C = 10^{-5} F, R_1 = 20 \Omega, R_2 = 10 \Omega $) as the coefficient $A$ of the constant phase element is varied. b)  Zoomed-in version of the phase portraits in  the first row, highlighting the visual competition between the features of the constant phase element and the capacitor for $A = 10^{-1}$ and $A = 10^{-3}$. For $A = 10^{-7}$, the constant phase element is negligible and only the capacitive features remain along the positive imaginary axis.}
    \label{fig:vary_A_CPE}
\end{figure}

The poles and zeros of the capacitive element compete with that of the constant phase element and can dominate them for some values of $A$. This is seen visually in  Fig. \ref{fig:vary_A_CPE}. For $A = 10^{-7}$, the constant phase element has negligible impact on the impedance measurements. If the impedance measurements are too few or too noisy, the rational approximation to the data might fail to detect the presence of a constant phase element altogether. Even if we have sufficient measurements with low noise, it is not trivial to separate the pole-zero structure of the constant phase element when applying the log substitution approach. Since $\alpha = \frac{1}{2}$ for the constant phase element in Equation \ref{eq:randles}, we expect to see poles or zeros at $2 \pi$ along the imaginary direction in the transformed phase portraits. 

\begin{figure}
    \centering
    \includegraphics[width=\textwidth]{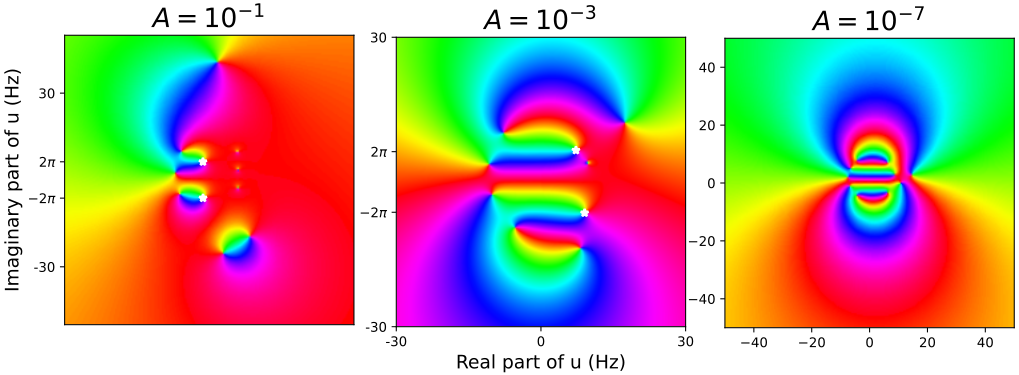}
    \caption{ Transformations of the phase portraits in Figure \ref{fig:vary_A_CPE} after using the log substitution. The zeros corresponding to those of  the constant phase element are highlighted by white stars. Without knowing the phase of the constant phase element, it would not be possible to disentangle the zeros.}
    \label{fig:vary_A_CPE_transformed}
\end{figure}

Indeed, we do see zeros at those locations (highlighted by white stars in Fig. \ref{fig:vary_A_CPE_transformed}). However, we are only able to disentangle those zeros as zeros of the constant phase element because we knew $\alpha  = \frac{1}{2} $ a priori.
\begin{figure}
    \centering    \includegraphics[width=\textwidth]{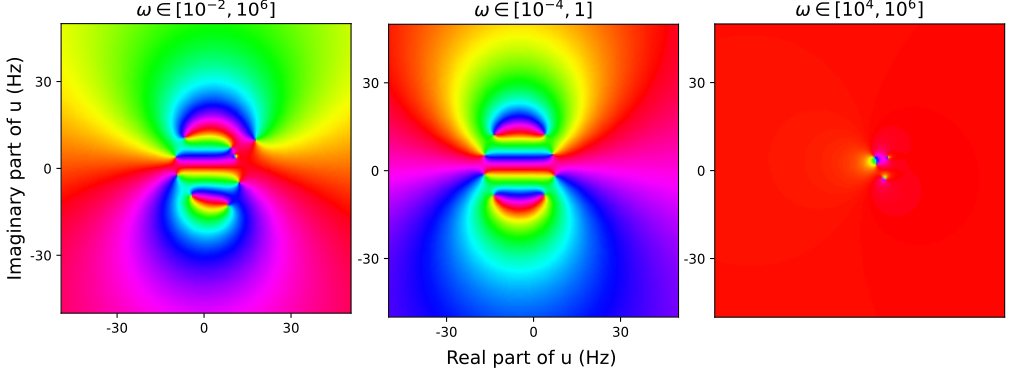}
    \caption{Transformed phase portraits for the Randles circuit (Equation \ref{eq:randles}, $A = 10^{-3} F, C = 10^{-5} F, R_1 = 20 \Omega, R_2 = 10 \Omega $) where the impedance has been sampled in different frequency regimes. (Left) The standard frequency range (200 logarithmically sampled measurements between 0.01 Hz and $10^6$ Hz). (Middle) The low frequency regime (200 logarithmically sampled measurements between 0.0001 Hz and 1 Hz). (Right) The high frequency regime (200 logarithmically sampled measurements between $10^4$ Hz and $10^6$ Hz).}
    \label{fig:prony}
\end{figure}

When such competing elements are present, it is worthwhile to collect and look at impedance measurements in particular frequency regimes where it is possible to isolate one or more elements. In Fig \ref{fig:prony}, we see that the transformed phase portraits look noticeably different. In the high frequency regime, the phase portrait appears to be nearly constant everywhere, indicating the impedance achieves a constant value at high frequency. The low frequency phase portrait, on the other hand,  displays structure similar to that of an exponential function (Fig. 5 of the main text).  To understand this structure, we examine the Randles circuit in the low frequency regime.

\begin{align}
Z_{low}( \omega) =\lim_{\omega \to 0} Z(\omega) = \lim_{\omega \to 0} R_1 + \frac{1}{C i \omega + \frac{1}{R_2 + 1/A\sqrt{i \omega}}} =   \frac{1}{C i \omega + A\sqrt{i \omega}}
    \label{eq:randles_limit}
\end{align}

Applying our substitution $u = \log(i\omega)$ , we can obtain

\begin{align}
\frac{1}{ Z_{low}(u)} =  C \exp(u) + A\exp(u/2)
    \label{eq:randles_limit_transformed}
\end{align}

Since we have a sum of exponentials, we can employ a classical linear algebra technique, Prony's method \cite{prony1795essai, weiss1963prony}, to recover the exponents. Prony's method or in practice, the regularized Prony method \cite{beylkin2010approximation,wilber2022data}, allows us to recover the exponents $\alpha_k$  and the coefficients $c_k$  from measurements $f (x)$ of the form shown in Equation \ref{eq:prony}. Details about Prony's method and its implementation can be found in Appendix \ref{appendix:prony}.

\begin{align}
f(x) = \sum_k c_k \exp(\alpha_k x)
\label{eq:prony}
\end{align}

Applying Prony's method to $1/ Z_{low}(u)$, we recover the exponent values (0.4998, 0.991). The coefficients (C, A) can also be recovered enabling us to identify a capacitor with capacitance $C$ and a constant phase element with impedance $Z(\omega) = 1/ A\sqrt{i \omega}$ from the impedance data. 

Competing elements in impedance measurements can thus be identified by examining specific frequency regimes and applying techniques such as Prony's method. However, the impedance measurements are typically collected by sampling logarithmically over a very large frequency range. This can result in too few measurements in the desired frequency regime. The mathematical analysis discussed in this section can inform experimentalists on where to concentrate sampling in frequency space during a second round of measurements. 

In this section, we discussed how the impedance measurements after the log substitution, takes the form of a sum of exponential functions in certain frequency regimes. In this setting, Prony's method allows us to identify the phase of the contributing components. However, since Prony’s method applies only in this restricted setting, there is strong motivation for exploring extensions of Prony's method \cite{potts2010parameter, stampfer2020generalized, wilber2022data, derevianko2023esprit,beylkin2010approximation} and further algorithmic developments, which would allow us to de-convolve overlapping features in more general, less idealized settings.

\section{ Analysis of impedance response measurements from a lithium-ion battery  coin cell.} \label{appendix:data}

In this section, we elaborate on some of  the mathematical analysis done for the system studied in Section 6 of the main text.

\subsection{ Deriving the low-frequency model.}
The rational approximation of the low-frequency impedance data reveals a simple structure shown in Figure 12c of the main text. We see a pole and zero lying very close to the positive imaginary axis. Since the pole is essentially at zero, the simplest model that shares these (approximate) features is a resistor and capacitor in series.

This low-frequency circuit model can be expressed by Equation \ref{eq:R_C_low_sup}. We expect the capacitive element to be imperfect with the exponent $\alpha$ since the zero does not lie exactly on the positive imaginary axis. 
\begin{align}
Z(\omega) =  R + \frac{1}{C (i \omega)^\alpha}  
    \label{eq:R_C_low_sup}
\end{align}
Because of the form of Equation \ref{eq:R_C_low_sup}, Prony's method can be applied to the low-frequency impedance data after using the transformation $u= \log(i \omega)$. Prony's method returns an exponent of 0 corresponding to  a purely resistive element and another exponent near 0.9 corresponding to a slightly imperfect capacitor, validating our model and allowing us to have a value for $\alpha$. 

 \begin{figure}
     \centering
    \includegraphics[width=\textwidth]{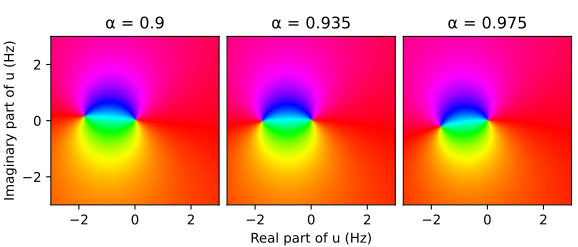}
\caption{ In order to find the exponent $\alpha$ for the imperfect capacitor, we make the substitution $u = (i \omega)^\alpha$. For the right value of $\alpha$, the poles and zeros must lie on the negative real axis (because the resistance $R$ and the capacitance $C$ are positive real numbers). This gives us a way to fine-tune $\alpha$ near the value of 0.9 (recovered by Prony's method).}
    \label{fig:lopez_low_pp}
\end{figure}

We can further fine-tune $\alpha$  (Figure \ref{fig:lopez_low_pp}) by using the substitution $u= (i \omega)^\alpha$ for different values of $\alpha$ so that the zero lies on the negative real axis (corresponding to an R-C circuit under the transformation). We recover an $\alpha$ of 0.935 and simultaneously $R = 108 \Omega, C = 5.3 \times 10^{-3} F$  from  the rational approximation. We thus are able to derive a complete model for the impedance data in the low-frequency regime.

\subsection{Deducing $Q$ and $\alpha$ from the low-frequency model.}

 This complete low-frequency model can be used to gain information about the unknowns $R_{ion}, Q, \alpha$ and $M(\omega)$ in Equation 16 from the main text. In the low frequency regime, Equation 17 in the main text becomes becomes Equation \ref{eq:transmission+M_low} written below.

  \begin{align}
\lim_{\omega \to 0} Z(\omega) = \frac{R_{ion}}{3} + \frac{1}{Q (i \omega)^\alpha} + \lim_{\omega \to 0}  M(\omega)
    \label{eq:transmission+M_low}
\end{align}

Since Prony's method returned only one non-zero exponent in the low frequency regime, we can safely assume that there is only one non-resistive element in the low-frequency regime. This implies that $Q$ in Equation  \ref{eq:transmission+M_low} must match $C$ in Equation \ref{eq:R_C_low_sup}. Likewise, $\alpha$ in Equation \ref{eq:transmission+M_low} must be 0.935. However, $R_{ion}$ cannot be similarly deduced because $M(\omega)$ could have resistive contributions in the low-frequency regime.

\subsection{Model parameters.} \label{appendix:parameters}

In Section 6 of the main text, we derive the following complete mathematical model.

 \begin{align}
    Z(\omega) =\frac{a(i \omega)^{\beta}+b}{(i \omega)^{\beta}+c}+  \sqrt{\frac{R_{ion}}{Q(i\omega)^\alpha}}\coth(\sqrt{R_{ion}Q(i\omega)^\alpha}) 
    \label{eq:complete_model_appendix}
\end{align}

From the low frequency model, we know that $Q = 5.3 \times 10^{-3} F$ and $\alpha = 0.935$. There are two minimal circuits shown in Figure 14 of the main text which are mathematically indistinguishable and yield  Equation \ref{eq:complete_model_appendix}.

\begin{table}[h!]
\centering
\caption{Circuit parameters extracted from the rational approximation for the choice of  $R_{ion}=90 \Omega$  and $R_{ion}=100 \Omega$. In Figure 15 of the main text, we see these parameters provide an excellent match to the impedance data.}
\label{tab:circuit_params}
\begin{tabular}{llccc}
\toprule
\textbf{$R_{ion}$ [$\Omega$]} & \textbf{Circuit} & \textbf{$R_1$ [$\Omega$]} & \textbf{$R_2$ [$\Omega$]} & \textbf{Impedance of imperfect capacitor $Z(\omega)$ [$\Omega$]} \\
\midrule
\multirow{2}{*}{90}  
 & a & $34.0$  & $41.7$  & $\dfrac{1}{1.1 \times 10^{-4}(i\omega)^{0.82}}$ \\
 & b & $75.7$  & $61.7$  & $\dfrac{1}{3.3 \times 10^{-5}(i\omega)^{0.82}}$ \\
\midrule
\multirow{2}{*}{100} 
 & a & $34.4$  & $40.35$ & $\dfrac{1}{1.06 \times 10^{-4}(i\omega)^{0.833}}$ \\
 & b & $74.75$ & $63.73$ & $\dfrac{1}{3.086 \times 10^{-5}(i\omega)^{0.833}}$ \\
\bottomrule
\end{tabular}
\end{table}

\end{document}